\documentclass[10pt,twocolumn]{IEEEtran}
\usepackage{setspace}
\usepackage{amssymb}
\usepackage{amsfonts}
\usepackage{amsmath, amsthm, amsfonts, amssymb, amsbsy}
\usepackage{amssymb}
\usepackage{setspace}
\usepackage{graphicx}
\usepackage{subfigure}
\usepackage{pstricks,arydshln}
\usepackage{multirow}
\usepackage{stfloats}
\usepackage{amscd,textcomp,amssymb,epsfig,graphics,epsf,color,amsmath,balance,algorithm,cite}
\usepackage{color}
\usepackage{bm}
\usepackage{epstopdf}
\usepackage[mathscr]{eucal}
\usepackage{pifont}
\usepackage{algpseudocode}
\usepackage{float}
\usepackage{hyperref}
\hypersetup{colorlinks=true,linkcolor=black,citecolor=black}
\usepackage{cite}
\graphicspath{{./src/img/}}

\def\BibTeX{{\rm B\kern-.05em{\sc i\kern-.025em b}\kern-.08em
    T\kern-.1667em\lower.7ex\hbox{E}\kern-.125emX}}
\usepackage{balance}

\newtheorem{theorem}{Theorem}
\theoremstyle{plain}\newtheorem{lemma}{Lemma}
\theoremstyle{plain}

\newtheorem{remark}{Remark}

\newtheorem{corollary}{Corollary}

\allowdisplaybreaks
\newcounter{TempEqCnt}
\begin{document}
\title{Wideband Beamforming for Near-Field Communications with Circular Arrays
\author{\large Yunhui Guo, \IEEEmembership{Student Member, IEEE}, Yang Zhang, \IEEEmembership{Member, IEEE}, \\
Zhaolin Wang, \IEEEmembership{Graduate Student Member, IEEE}, and Yuanwei Liu, \IEEEmembership{Fellow, IEEE}}\vspace{-1.5mm}
\thanks{Y. Guo is with the State Key Laboratory of Integrated Service Networks, Xidian University, Xi'an 710071, China, and also with the School of Electronic Engineering and Computer Science, Queen Mary University of London, London E1 4NS, U.K. (Email: yunhuiguo@stu.xidian.edu.cn).}
\thanks{Y. Zhang is with the State Key Laboratory of Integrated Service Networks, Xidian University, Xi'an 710071, China (Email: yangzhang1984@gmail.com).}
\thanks{Zhaolin Wang and Yuanwei Liu are with the School of Electronic Engineering and Computer Science, Queen Mary University of London, London E1 4NS, U.K. (e-mail: zhaolin.wang@qmul.ac.uk, yuanwei.liu@qmul.ac.uk).}
}

\maketitle \thispagestyle{empty}
\begin{abstract}
The beamforming performance of the uniform circular array (UCA) in near-field wideband communication systems is investigated. Compared to uniform linear array (ULA), UCA exhibits uniform effective array aperture in all directions, thus enabling more users to benefit from near-field communications. In this paper, the unique beam squint effect in near-field wideband UCA systems is comprehensively analyzed in both the distance and angular domains. It is rigorously demonstrated that the beam focal point only exists at a specific frequency in wideband UCA systems, resulting in significant beamforming loss. To alleviate this unique beam squint effect, the true-time delay (TTD)-based beamforming architecture is exploited. In particular, two wideband beamforming optimization approaches leveraging TTD units are proposed. 1) \emph{Analytical approach}: In this approach, the phase shifters (PSs) and the time delay of TTD units are designed based on the analytical formula for beamforming gain. Following this design, the minimum number of TTD units required to achieve a predetermined beamforming gain is quantified. 2) \emph{Joint-optimization approach}: In this method, the PSs and the TTD units are jointly optimized under practical maximum delay constraints to approximate the optimal unconstrained analog beamformer. Specifically, an efficient alternating optimization algorithm is proposed, where the PSs and the TTD units are alternately updated using either the closed-form solution or the low-complexity linear search approach. Extensive numerical results demonstrate that 1) the proposed beamforming schemes effectively mitigate the beam squint effect, and 2) the joint-optimization approach outperforms the analytical approach in terms of array gain and achievable spectral efficiency.
\end{abstract}

\vspace{-1mm}

\vspace{-2mm}
\section{I\footnotesize{NTRODUCTION}}\label{sectionI}
Beyond fifth generation (B5G) and sixth generation (6G) communication networks are expected to facilitate numerous high data rate applications, including virtual reality/augmented reality (VR/AR), digital twins, and holographic telepresence \cite{1}. To fulfill the requirements of these 6G applications, extremely large-scale antenna arrays (XLAAs) and high-frequency are considered as promising technologies \cite{2}. Firstly, benefiting from the enormous spectral resources available in the millimeter-wave (mmWave) and terahertz (THz) frequency band, high-frequency communications can provide tens of gigahertz (GHz) bandwidth \cite{3,4}. Furthermore, in contrast to 5G massive multiple-input multiple-output (MIMO) systems, extremely large-scale MIMO (XL-MIMO) is equipped with ten times more antennas at the base station (BS), significantly enhancing beamforming gain and achieving higher spectral efficiency \cite{5}.

Nonetheless, the emergence of high frequencies and XLAAs may introduce the fundamental change in the electromagnetic (EM) propagation environment \cite{6, 7}. It is well-established that the EM radiation field can be divided into near-field and far-field regions. The boundary between these two regions is commonly distinguished by the Rayleigh distance, expressed as ${d_r} = \frac{{2{R^2}}}{\lambda_c }$, where ${R}$ represents the antenna aperture and $\lambda_c$ denotes the wavelength \cite{8}. For conventional massive MIMO communication systems, the EM propagation model can be simplified as the far-field planar-wave model due to the relatively small scale of the BS antenna array. However, with higher frequency and dramatically increased array aperture, users are more likely to be located in the near-field region. In this case, the widely adopted far-field planar-wave model becomes inaccurate since the non-negligible phase discrepancy between the spherical wave and planar-wave models \cite{9}. Consequently, it is necessary to redesign wireless networks by considering \hspace{-0.2mm}the \hspace{-0.2mm}spherical \hspace{-0.1mm}wave characteristics of the near-field.

\subsection{Prior Works}
On the one hand, the fundamental changes in EM propagation characteristics in the near-field region provide new opportunities for enhancing the performance of 6G wireless communication systems. Specifically, in far-field communication, the signal wavefront is commonly approximated as planar, enabling the far-field beamforming to steer the beam towards specific user direction. This beamforming method is referred to as far-field beamsteering. Nevertheless, in near-field communication, near-field beamforming generates angle-distance correlated beams that are focused around the user locations. This is commonly known as near-field beamfocusing \cite{10}. Compared with far-field beamsteering, near-field beamfocusing considers both the distance and the direction between the transmitter and users \cite{11}. This can provide additional degrees of freedom (DoFs) to mitigate multi-user interference and improve system performance \cite{12,13,14}. To be specific, reference \cite{12} demonstrates that evanescent waves can significantly improve the spatial DoF and increase capacity within the reactive near-field region. To further harness the increased DoFs to enhance spectral efficiency, the continuous aperture MIMO that multiplex data streams via different transmission modes is proposed in \cite{13}. Additionally, \cite{14} highlights that distance information functions as an exploitable and valuable dimension in multi-user communications, allowing for effectively suppressing multi-user interference.

On the other hand, the implementation of XLAAs in high-frequency systems presents significant challenges attributed to their elevated hardware costs and high power consumption\cite{15}. Specifically, the conventional fully digital architectures, where each transmit antenna attaches to a dedicated radio frequency (RF) chain, are impractical for XLAAs systems. To achieve a balance between cost and performance, hybrid analog and digital architectures are considered to be adopted. In contrast to conventional digital beamforming schemes, hybrid beamforming is more efficient as it avoids the use of numerous higher hardware costs and power-consuming RF chains \cite{16}. This is achieved by dividing a large-dimensional digital beamformer into a small-dimensional digital beamformer with RF chains and a large-dimensional analog beamformer realized through phase shifters (PSs) \cite{17}. However, in wideband XL-MIMO systems, PSs can only generate the same phase shifts at different frequencies, known as frequency-independent \cite{18}. In the context of the orthogonal frequency-division multiplexing (OFDM) data transmission format, the utilization of frequency-independent PSs leads to beams at different subcarrier frequencies pointing towards distinct physical directions, resulting in spatial direction deviation for each subcarrier. This phenomenon is commonly known as the beam squint effect, also referred to as the spatial-frequency wideband effect\cite{19}. It is important to mention the beam squint effect leads to severe beamforming loss, significantly reducing the achievable rate.

To effectively mitigate the beam squint effect in near-field wideband communication systems, recent researches \cite{20,21,22} have incorporated true-time delay (TTD) units, which introduce specific time delays to signals, thereby creating frequency-dependent phase shifts. In particular, reference \cite{20} proposes a phase-delay focusing (PDF) method for near-field beamfocusing, where a piecewise-far-field approximation is proposed for wideband beamforming design. In \cite{21}, a novel sub-connected TTD-based beamforming architecture is investigated to alleviate the near-field wideband beam squint effect and facilitate the beamfocusing. Furthermore, \cite{22} develops a low-complexity beamforming method for the TTD-PS \hspace{-2mm}architecture, \hspace{-0.2mm}enabling \hspace{-0.3mm}the \hspace{-0.3mm}learning \hspace{-0.3mm}and \hspace{-0.3mm}optimization \hspace{-0.4mm}of \hspace{-0.2mm}near-field wideband beams without requiring any channel knowledge.

Based on the aforementioned research, the TTD-based beamforming architectures prove to be a promising solution for designing wideband beamforming in near-field communications. Nevertheless, the existing research in this area has primarily concentrated on the deployment of uniform linear array (ULA) \cite{18,19,20,21,22}. It is worth highlighting that the effective array aperture dramatically reduces at large incident/departure angles for the widely used ULA \cite{23}. This leads to severe distortion in the beampattern near the edges of ULA. Meanwhile, since the near-field region is affected by the effective array aperture, reduced array aperture for ULA results in reduced near-field region \cite{24}.

To tackle this issue, the uniform circular array (UCA) with rotational symmetry property is considered to ensure a large and uniform effective array aperture at any angle. Some preliminary studies have been made to investigate the utilization of UCA for enhancing system performance in near-field communications \cite{25,26,27}. For instance, the authors of \cite{25} demonstrate that UCA exhibits the capability of providing a wider scanning range and invariant beampattern along the entire azimuth plane, contributing to enhanced stability and performance of near-field communications. Similarly, \cite{26} indicates that UCA outperforms uniform rectangular array (URA) and ULA in terms of achieving higher spectral efficiency performance. Reference \cite{27} illustrates that employing UCA allows for deploying a greater number of antennas within a limited physical space. This results in higher antenna array gain, effectively overcoming propagation attenuation in near-field communications. However, the existing research has rarely focused the beamforming design for wideband UCA systems in near-field communications.

\subsection{Motivation and Contributions}
Against the above background, the beamforming design for wideband UCA systems may introduce the beam squint effect, i.e., the mismatch between the frequency-independent phase shifts of PSs and required frequency-dependent phase shifts to perform beamforming. To the best of the authors' knowledge, the beamforming performance of UCA in near-field wideband communication systems has not been explored in existing research, and there are no practical solutions for beamforming design to overcome this fundamental challenge. Motivated by this, we first investigate the beamforming performance of UCA in near-field wideband communication systems, and then analyze the performance loss caused by the beam squint effect. To effectively mitigate the near-field beam squint effect in wideband UCA systems, two wideband beamfocusing optimization approaches leveraging TTD units are proposed. The main contributions can be summarized as follows.
\begin{itemize}
    \item We first characterize the near-field beam squint effect for wideband UCA systems in both the distance and angular domains. The analysis results validate that the beamforming gain can be approximated using the zero-order Bessel function in both the distance and angular domains. Furthermore, it is also demonstrated that the beam focal point only exists at a specific frequency in wideband UCA systems, resulting in significant beamforming loss.
    \item We propose an analytical method for near-field beamforming design. This approach exploits the TTD-based beamforming architecture, effectively \hspace{-0.3mm}mitigating \hspace{-0.3mm}the \hspace{-0.3mm}beam \hspace{-0.3mm}squint effect. The basic idea of the analytical approach is to design the PSs and the time delay of TTD units based on the analytical formula for beamforming gain. Subsequently, we quantify the minimum number of TTD units required to produce a predefined array gain performance for wideband UCA systems in near-field communications.
    \item We further propose a low-complexity joint-optimization near-field beamforming design method. In this approach, both the TTD and PS beamformers are alternately optimized under practical maximum delay constraints to approximate the optimal unconstrained analog beamformer. Specifically, the optimization variables are iteratively optimized by closed-form solutions or low-complexity one-dimensional search.
    \item Extensive numerical results are provided to demonstrate the effectiveness of the proposed beamforming optimization methods, revealing that 1) the precise beamforming design in both the distance and angular domains is essential for wideband UCA systems in near-field communications; 2) the beam squint effect can be effectively mitigated by utilizing the TTD-based hybrid beamforming architectures; 3) the joint-optimization approach achieves higher spectral efficiency than the analytical approach for different bandwidths and TTD unit maximum delays.
\end{itemize}

\subsection{Organization and Notations}
The remaining of this paper is organized as follows. The wideband UCA system model for near-field communications is presented in Section \ref{sectionII}. Section \ref{sectionIII} characterizes the beamforming properties of wideband UCA systems and reveals the near-field beam squint effect in both the distance and angular domains. Section \ref{sectionIV} presents the proposed analytical beamforming algorithm. Section \ref{sectionV} describes the joint-optimization beamforming algorithm. The numerical results are illustrated in Section \ref{sectionVI}. Finally, we conclude the paper in Section \ref{sectionVII}.

\emph{Notations}: Vectors and matrices are represented by the boldface lower-case symbols and capital boldface characters. The operators ${\left( \cdot \right)^T}$, ${\left( \cdot \right)^{-1}}$, and ${\left( \cdot \right)^H}$ represent the transpose, inverse, and Hermitian transpose of a matrix or vector, respectively. $\|\cdot\|$ and $\left|  \cdot  \right|$ denote the Euclidean norm and scalar norms, respectively. ${\left[ {\bf{x}} \right]_n}$ stands for the $n$-th element of the vector ${\bf{x}}$. A block diagonal matrix with diagonal blocks ${{{\bf{x}}_1}, \cdots  ,{{\bf{x}}_N}}$ can be denoted as ${\rm{blkdiag}}\left( {{{\bf{x}}_1}, \cdots ,{{\bf{x}}_N}} \right)$. Moreover, $\mathcal{CN}\left( {0,{\sigma ^2}} \right)$ represents a complex Gaussian distribution with mean 0 and variance ${{\sigma ^2}}$. Finally, we use ${\rm{Re}}\left\{  \cdot  \right\}$ to represent the real component of a complex, and the symbol $\angle$ to represent the phase of a complex value.

\section{\normalsize{S}\footnotesize{YSTEM} \normalsize{M}\footnotesize{ODEL}}\label{sectionII} 
\begin{figure}[t]
\centering
\vspace{-0.5mm}
\includegraphics[width=0.39\textwidth]{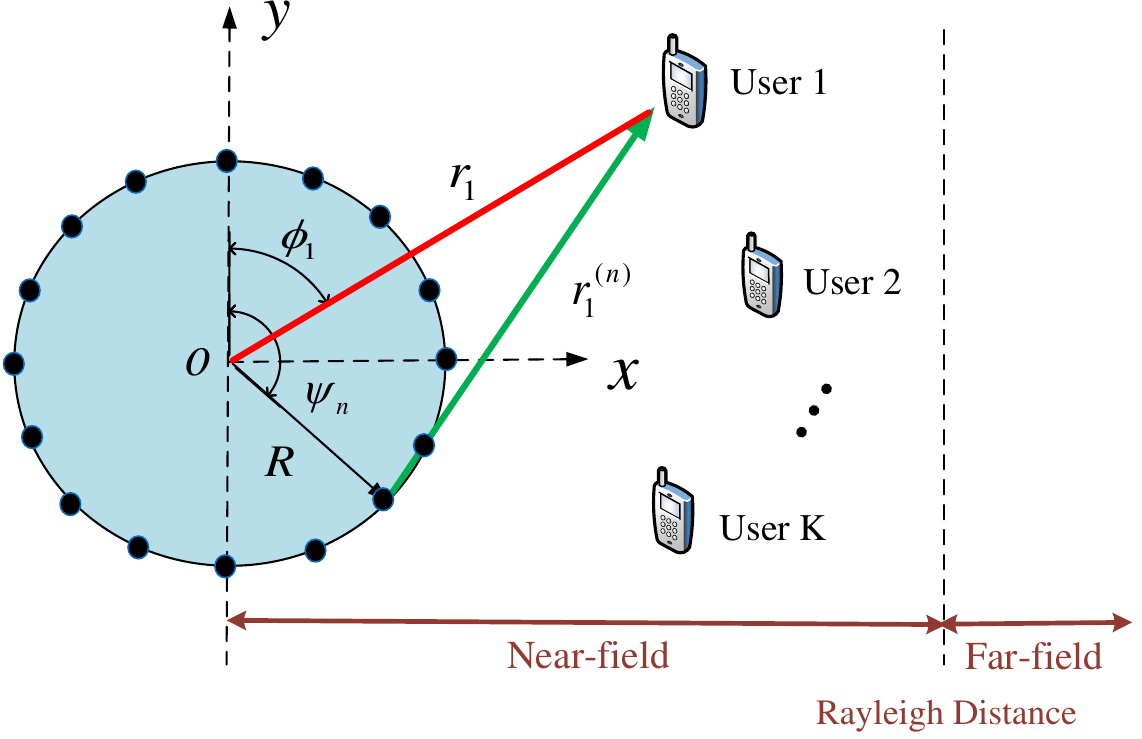}\vspace{-2.5mm}
\caption{Illustration of the near-field channel model.}\vspace{-2mm}
\label{fig1}
\end{figure}

\vspace{-0.5mm}
We investigate a near-field wideband multi-user communication system, where a BS employs an $N$-element UCA to serve $K$ single-antenna users using OFDM with $M$ subcarriers. Let $B$ and $f_c$ denote the signal bandwidth and carrier frequency, respectively. Subsequently, the $m$-th subcarrier frequency can be determined as $\tilde{f}_m = -\frac{B}{2}+\frac{(m-1) B}{M-1}, m \in \mathcal{M} \triangleq\{1,\cdots , M\}$. Moreover, according to the UCA geometry structure, the antennas are uniformly positioned along the circle with a radius of $R$, as depicted in Fig. \ref{fig1}. To simplify the description, the polar coordinate is utilized to express the location of the $n$-th transmit antenna element, denoted by ${{\bf{p}}_n} = \left( {R,{\psi _n}} \right)$, where ${\psi _n} = \frac{{2\pi n}}{N}$ for $n = 0, \cdots ,N-1$. Then, we assume that users are restricted to the two-dimensional space of the same plane as the UCA. Utilizing polar coordinates, the position of user $k$ is represented as $\left( {{r_k},\!{\phi _k}} \right)$, where $r_k$ signifies the distance between the transmitter and the user $k$ and ${\phi _k}$ denotes the elevation angles of departure.

Considering the significant path loss caused by the scatters, mmWave/THz communications primarily rely on the line-of-sight (LOS) propagation path. As a result, our primary focus is on the LOS channel. Specifically, we model the LOS channel based on the EM propagation characteristics in the radiating near-field, i.e., the phase of each transmit antenna is modeled using its exact distance from the user. Thus, the near-field channel ${h_{m,k,n}}$ between the user $k$ and the $n$-th BS antenna at the $m$-th subcarrier could be described as\vspace{-0.5mm}
\begin{equation}
\label{1}
{h_{m,k,n}} = g_{m,k}^{(n)}{e^{ - j{k_m}r_k^{(n)}}},\vspace{-1mm}
\end{equation}
where $k_m = \frac{{2\pi {f_m}}}{c}$. ${r_k^{(n)}}$ represents the propagation distance between the $n$-th transmit antenna element and the user $k$, which can be constructed as\vspace{-0.5mm}
\begin{align}
\label{2}
r_k^{(n)} &= \sqrt {{r_k}^2 + {R^2} - 2{r_k}R\cos \left( {\phi_k  - {\psi _n}} \right)}\nonumber \\
&\hspace{-0.2mm}\mathop \approx \limits^{(a)} {r_k} - R\cos \left( {\phi_k  - {\psi _n}} \right) + \frac{{{R^2}}}{{2r_k}}\left( {1 - {{\cos }^2}\left( {\phi_k  - {\psi _n}} \right)} \right)\nonumber\\
&\mathop  = {r_k} + \varsigma _{{r_k},\phi_k }^{(n)},
\end{align}
where $\varsigma _{{r_k},\phi_k }^{(n)} =  - R\cos \left( {{\phi _k} - {\psi _n}} \right) + \frac{{{R^2}}}{{2{r_k}}}\left( {1 - {{\cos }^2}\left( {{\phi _k} - {\psi _n}} \right)} \right)$ denotes the difference between the propagation distance to the $n$-th antenna and the distance to the origin. The approximation ${\left( a \right)}$ is derived from the second-order Taylor series expansion $\sqrt{1+x}=1+\frac{x}{2}-\frac{x^2}{8}+\mathcal{O}\left(x^3\right)$. This approximation is sufficiently accurate when propagation distance $r_k$ is larger than the Fresnel distance, which is typically comparable to the array aperture and thus could be commonly satisfied \cite{8}.

Furthermore, according to \cite{28}, the specific free-space path gain $g _{m,k}^{(n)}$ is given by\vspace{-2.5mm}
\begin{equation}
\label{3}
g_{m,k}^{(n)} = \frac{{{\lambda _m}}}{{4\pi r_k^{(n)}}},\vspace{-1 mm}
\end{equation}
where $\lambda_m=\frac{c}{f_m}$ stands for the wavelength at $f_m$ and $c$ is the light speed. Generally, the distance ${r_k^{(n)}}$ between the transmitter and user $k$ is larger than the antenna array aperture, i.e., ${r_k} \gg R$, we can assume $g _{m,k}^{(0)} \approx  \cdots  \approx g _{m,k}^{(N-1)} \approx {g_{m,k}} = \frac{{{\lambda _m}}}{{4\pi {r_k}}}$ based on the Fresnel approximation \cite{29}. Consequently, the near-field LOS channel vector of the $k$-th user ${\bf{h}}\left( {{r_k},{\phi _k},{f_m}} \right)\in {\mathbb{C}^{N \times 1}}$ can be further represented by
\begin{align}
\label{4}
{\left[ {{\bf{h}}\left( {{r_k},{\phi _k},{f_m}} \right)} \right]_n} & = {g_{m,k}}{\!\left[ {{e^{ - j{k_m}r_k^{(0)}}}\!,\! \cdots \!,{e^{ - j{k_m}r_k^{(N - 1)}}}} \right]^T}\nonumber\\
& = {\sqrt N} {g_{m,k}}{{\bf{a}}_m}({r_k},{\phi _k}),
\end{align}
where ${{\bf{a}}_m}(r_k,\phi_k)$ represents the near-field array response vector. Based on equation \eqref{4}, we can further approximate the $n$-th element of ${{\bf{a}}_m}(r_k,\phi_k)$ as follows \vspace{-1mm}
\begin{align}
\label{5}
\hspace{-0.8mm}{\left[ {{{\bf{a}}_m}\!({r_k},{\phi _k})} \right]_n} &\!\!\approx\! \frac{1}{{\sqrt N }}{e^{ - j{k_m}\left( \!{{r_k} - R\cos \left( {{\phi _k} - {\psi _n}} \right) + \frac{{{R^2}}}{{2{r_k}}}\left( {{{\sin }^2}\left( {{\phi _k} - {\psi _n}} \right)}\! \right)} \!\right)}}\nonumber\\
&\!\!=\! {e^{ - j{k_m}{r_k}}}{\left[ {{{\bf{b}}_m}({r_k},\phi_k )} \right]_n},
\end{align}
where ${\left[ {{{\bf{b}}_m}\!({r_k},\!{\phi _k})} \right]_n} \!\!= \!\! \frac{1}{\sqrt N}{e^{j{k_m}\!\left( \!{R\cos \left( {{\phi _k} - {\psi _n}} \right) - \frac{{{R^2}}}{{2{r_k}}}\left( {{{\sin }^2}\left( {{\phi _k} - {\psi _n}} \right)} \!\right)} \!\right)}}$ represents the $n$-th element of vector ${{{\bf{b}}_m}({r_k},\phi_k )}$. Since the phase ${e^{ - j{k_m}{r_k}}}$ in \eqref{5} is independent of the antenna index $n$, our primary focus is to discuss the vector ${{{\bf{b}}_m}({r_k},\phi_k )}$. Furthermore, it can be observed that the near-field beamfocusing vectors \eqref{5} can concentrate power on specific angles and distances, i.e., specific locations in the whole two-dimensional space. This is significantly different from the far-field beamsteering vectors, which can only focus the signal power on specific angles \cite{20}. Consequently, in a mmWave/THz XL-MIMO system, apart from the angle informations of users, the distance informations also play an essential role when performing beamforming. In the subsequent section, we will conduct a more in-depth analysis to characterize the beamforming properties of wideband UCA systems in both distance and angular domains.

\begin{remark} \label{remark_1}
    \emph{Note that when Taylor series expansion retains only the first-order term, the distance $ r_k^{(n)}$ simplifies to the far-field condition, i.e., $r_k^{(n)} \mathop  \approx {r_k} - R\cos \left( {\phi_k - {\psi _n}} \right)$. In this case, the near-field beamfocusing vector naturally degenerates into the far-field form. This implies that the far-field beamsteering vector is essentially a special case of near-field beamfocusing vector without higher-order Taylor series expansions.}
\end{remark}

\section{\normalsize{A}\footnotesize{NALYSIS} OF THE \normalsize{N}\footnotesize{EAR}-\normalsize{F}\footnotesize{IELD} \normalsize{B}\footnotesize{EAM} \normalsize{S}\footnotesize{QUINT} \normalsize{E}\footnotesize{FFECT} IN \normalsize{W}\footnotesize{IDEBAND} \normalsize{UCA} \normalsize{S}\footnotesize{YSTEMS} }\label{sectionIII}
In this section, we first characterize the beamforming properties of wideband UCA systems in near-field communications and subsequently reveal the near-field beam squint effect. Specifically, the beamforming loss caused by the near-field beam \hspace{-0.4mm}squint \hspace{-0.4mm}effect \hspace{-0.5mm}is \hspace{-0.4mm}analyzed \hspace{-0.5mm}in angular and distance \hspace{-0.5mm}domains.

\subsection{Near-Field Beam Squint Effect}
In hybrid beamforming architectures, the digital beamformer and analog beamformer are integrated to generate directional and less interfered beams, thereby achieving high spatial multiplexing gain. In particular, analog beamforming aims to create high-gain beams aligned with the desired user directions by precisely adjusting the phase of the signals. To achieve this objective, PSs are introduced to compensate for phase differences between different antennas and form equal-phase planes. Utilizing the conventional PSs \cite{30}, the near-field beamforming vector ${{\bf{w}}_{{\rm{PS}}}}({r_k},{\phi _k}) \in {\mathbb{C}^{N \times 1}}$ is typically determined based on the channel at the center frequency $f_c$ to align with the desired user location $({r_k},{\phi_k})$, i.e.,\vspace{-2mm}
\begin{align}
\label{6}
\hspace{-2.3mm}{{\bf{w}}_{{\rm{PS}}}}({r_k},{\phi _k}) &\!= \!{\bf{b}}_c ({r_k},{\phi _k}) \nonumber\\
 &\! = \!\frac{1}{{\sqrt N }}{e^{j{k_c}\left( {R\cos \left( {{\phi _k} - {\psi _n}} \right) - \frac{{{R^2}}}{{2{r_k}}}\left( {{{\sin }^2}\left( {{\phi _k} - {\psi _n}} \right)} \right)} \right)}},
\end{align}
where ${\bf{b}}_c ({r_k},\phi_k)$ denotes the array response vector at center frequency $f_c$ and ${k_c} = \frac{{2\pi {f_c}}}{c}$. Notice that the phase ${k_c}\left( {R\cos \left( {{\phi _k} - {\psi _n}} \right) - \frac{{{R^2}}}{{2{r_k}}}\left( {{{\sin }^2}\left( {{\phi _k} - {\psi _n}} \right)} \right)} \right)$ in \eqref{6} is frequency-independent. However, the term $k_m = \frac{{2\pi {f_m}}}{c}$ in the array response vector in equation \eqref{5} is frequency-dependent, which mismatches the frequency-independent beamforming vector ${{\bf{w}}_{{\rm{PS}}}}({r_k},{\phi _k})$. This mismatch results in the beams generated by ${{\bf{w}}_{{\rm{PS}}}}$ at different frequencies will be focused on different locations, leading to significant beamforming gain loss. This phenomenon is commonly described as beam squint effect, also known as the spatial-frequency wideband effect.\vspace{-0.5mm}
\begin{remark} \label{remark_2}
    \emph{Note that the beam squint effect can be negligible in narrowband systems, as the phase shift in the beamsteering vector only exhibits limited frequency-dependent variation. Nevertheless, in wideband systems, the range of phase variations expands significantly, resulting in severe beamforming loss. A detailed analysis of this effect will be conducted in the following section.}
\end{remark}

\subsection{Analysis of Beamforming Gain in the Angular Domain}
Based on the design method of matched-filter beamforming \cite{31}, the beamforming vectors can be formulated by the conjugates of near-field beamfocusing vectors. Subsequently, the expression for the beamforming gain is given by\vspace{-1mm}
\begin{align}
\label{7}
\hspace{-1mm}{G_m}\!\!\left( {{r_1},{\phi _1},{r_2},{\phi _2}} \right)& \!\!= \!\left| {{\bf{b}}_m^H({r_1},{\phi _1}){{\bf{b}}_c}({r_2},{\phi _2})} \right| \nonumber\\
& \hspace{-1.5mm}= \!\! \frac{1}{N}\left|\! {\sum\limits_{n = 0}^{N - 1} {{e^{j{k_m}\varsigma _{{r_1},{\phi _1}}^{(n)}}}}  \times {e^{ - j{k_c}\varsigma _{{r_2},{\phi _2}}^{(n)}}}} \right|\nonumber\\
& \hspace{-1.6mm} =\!\! \frac{1}{N} \!\left| {\sum\limits_{n = 0}^{N - 1} \!{{e^{ j{k_m}\! \left(\! {\sqrt {{r_1}^2 + {R^2} - 2{r_1}R\cos \left( {{\phi _1} - {\psi _n}} \right)}  - {r_1}} \! \right)}}} } \right.\nonumber\\
&\hspace{+2mm} \left. { \times {e^{ - j{k_c}\left( {\sqrt {{r_2}^2 + {R^2} - 2{r_2}R\cos \left( {{\phi _2} - {\psi _n}} \right)}  - {r_2}} \right)}}} \right|.
\end{align}

Since we are examining the variation of beamforming gain with respect to the radiating angle, we assume a specific propagation distance, denoted as ${r_1} = {r_2} = r$. Then, the characteristics of beamforming gain can be described by the subsequent lemma.\vspace{-1mm}
\begin{lemma} \label{lemma_1}
    \emph{If the frequency-independent near-field beamsteering vector ${{\bf{b}}_c}(r,{\phi _1})$ is utilized, the attained beamforming gain at frequency $f_m$ at any direction $\phi_2$ can be formulated as}\vspace{-6mm}
\end{lemma}
\begin{align}
\label{8}
{G_m}\left( {r,\phi_1,r,{\phi_2}} \right) = \left| {{J_0}\left( \eta  \right)} \right|,
\end{align}
where ${J_0}\left(  \cdot  \right)$ denotes the zero-order Bessel function. The variable $\eta$ is defined as
\begin{align}
\label{9}
\eta  = R\sqrt {k_c^2 + k_m^2 - 2{k_c}{k_m}\cos \left( {\phi_1 - {\phi_2}} \right)},
\end{align}
where ${k_m} = \frac{{2\pi {f_m}}}{c}$ and ${k_c} = \frac{{2\pi {f_c}}}{c}$.
\begin{proof}
The detailed mathematical proof of \textbf{Lemma \ref{lemma_1}} can be found in Appendix \ref{proofA}.
\end{proof}

Based on the characteristics of ${J_0}\left(  \cdot  \right)$, it can be observed that ${J_0}\left( \eta  \right)$ decreases with the increase of $\eta$. Therefore, the Bessel function reaches its maximum value when satisfied $\eta$ = 0, which corresponds to the maximum beamforming gain with ${f_m} = {f_c}$ and $\phi_1 = {\phi_2}$. In other words, high-gain beams can only be achieved at central frequency $f_c$. This leads to a significant loss in array gain at non-central frequencies.

\begin{remark} \label{remark_3}
    \emph{Note that in the proof of \textbf{Lemma \ref{lemma_1}}, we only retain the first-order term of the Taylor series. This approximation is motivated by the fact that even in the near-field region, the first-order term remains dominant when users are at the same distance. The accuracy of our approximate analysis is further confirmed by the simulation results discussed in Section \ref{sectionVIA}.}
\end{remark}

\subsection{Analysis of Beamforming Gain in the Distance Domain}
Subsequently, we concentrate on the beamforming performance in the distance domain, where ${\phi _1} = {\phi _2} = \phi $ is assumed. The expression for the beamforming gain is given by\vspace{-1mm}
\begin{align}
\label{10}
        {G_m}\left( {{r_1},\phi ,{r_2},\phi } \right) & = \left| {{\bf{b}}_m^H({r_1},\phi ){{\bf{b}}_c}({r_2},\phi )} \right| \nonumber\\
& \approx  \frac{1}{N}\left| \mathop \sum\limits_{n = 0}^{N - 1} {e^{j{k_m}\varsigma _{{r_1},\phi }^{(n)}}} \times {e^{ - j{k_c}\varsigma _{{r_2},\phi }^{(n)}}} \right| \nonumber\\
& = \frac{1}{N}\left| {\sum\limits_{n = 0}^{N - 1} {{e^{-jR\left( {{k_m} - {k_c}} \right)\cos \left( {\phi  - {\psi _n}} \right)}}} } \right. \nonumber\\
& \hspace{4mm} \left. { \times {e^{j{R^2}\left( {\frac{{{k_m}}}{{2{r_1}}} - \frac{{{k_c}}}{{2{r_2}}}} \right)\left( {1 - {{\cos }^2}\left( {\phi  - {\psi _n}} \right)} \right)}}} \right|,
\end{align}
where the definitions of ${\varsigma _{{r_1},\phi }^{(n)}}$ and ${\varsigma _{{r_2},\phi }^{(n)}}$ are described in equation \eqref{2}. Following this, the beamforming gain across different distances could be further approximated using the subsequent lemma.\vspace{-1.5mm}
\begin{lemma} \label{lemma_2}
    \emph{Utilizing frequency-independent phase shifts constructed based on the central frequency $f_c$, the attained beamforming gain with UCA at frequency $f_m$ can be represented by}\vspace{-8mm}
\end{lemma}
\begin{align}
\label{11}
{G_m}\left( {{r_1},\phi ,{r_2},\phi } \right) &= \left| {{\bf{b}}_m^H({r_1},\phi ){{\bf{b}}_c}({r_2},\phi )} \right|\nonumber\\
& \approx \left| {{J_0}\left( {R\left( {{k_c} - {k_m}} \right) + \varpi } \right)} \right|,
\end{align}
where $\varpi  = {R^2}\left( {\frac{{{k_c}}}{{4{r_2}}} - \frac{{{k_m}}}{{4{r_1}}}} \right)$.

\begin{proof}
The detailed mathematical proof of \textbf{Lemma \ref{lemma_2}} is provided in Appendix \ref{proofB}.
\end{proof}

To be specific, \textbf{Lemma \ref{lemma_2}} illustrates the characteristics of near-field beamfocusing in the distance domain. In contrast to far-field beamsteering, which remains uniform beamforming gain across different distances, the beamforming gain in the near-field fluctuates with distances. A more comprehensive analysis is provided in Section \ref{sectionVIA}.

\section{\normalsize{A}\footnotesize{LLEVIATING} \normalsize{B}\footnotesize{EAM} \normalsize{S}\footnotesize{QUINT} \normalsize{E}\footnotesize{FFECT} \normalsize{W}\footnotesize{ITH} \normalsize{TTD}-\normalsize{B}\footnotesize{ASED} \normalsize{B}\footnotesize{EAMFORMING} \normalsize{A}\footnotesize{RCHITECTURE} }\label{sectionIV}
As previously discussed, the beam squint effect in conventional PS-based beamforming architectures leads to severe beamforming loss. To alleviate the beamforming loss, the TTD-based beamforming architecture is utilized in this section. To be specific, the beamforming gain for wideband UCA systems will be investigated, and subsequently the analytical beamforming algorithm will be proposed.

\subsection{TTD-PS-based Beamforming Architecture}
\begin{figure}[t]
\centering
\vspace{-1mm}
\includegraphics[width=0.37\textwidth]{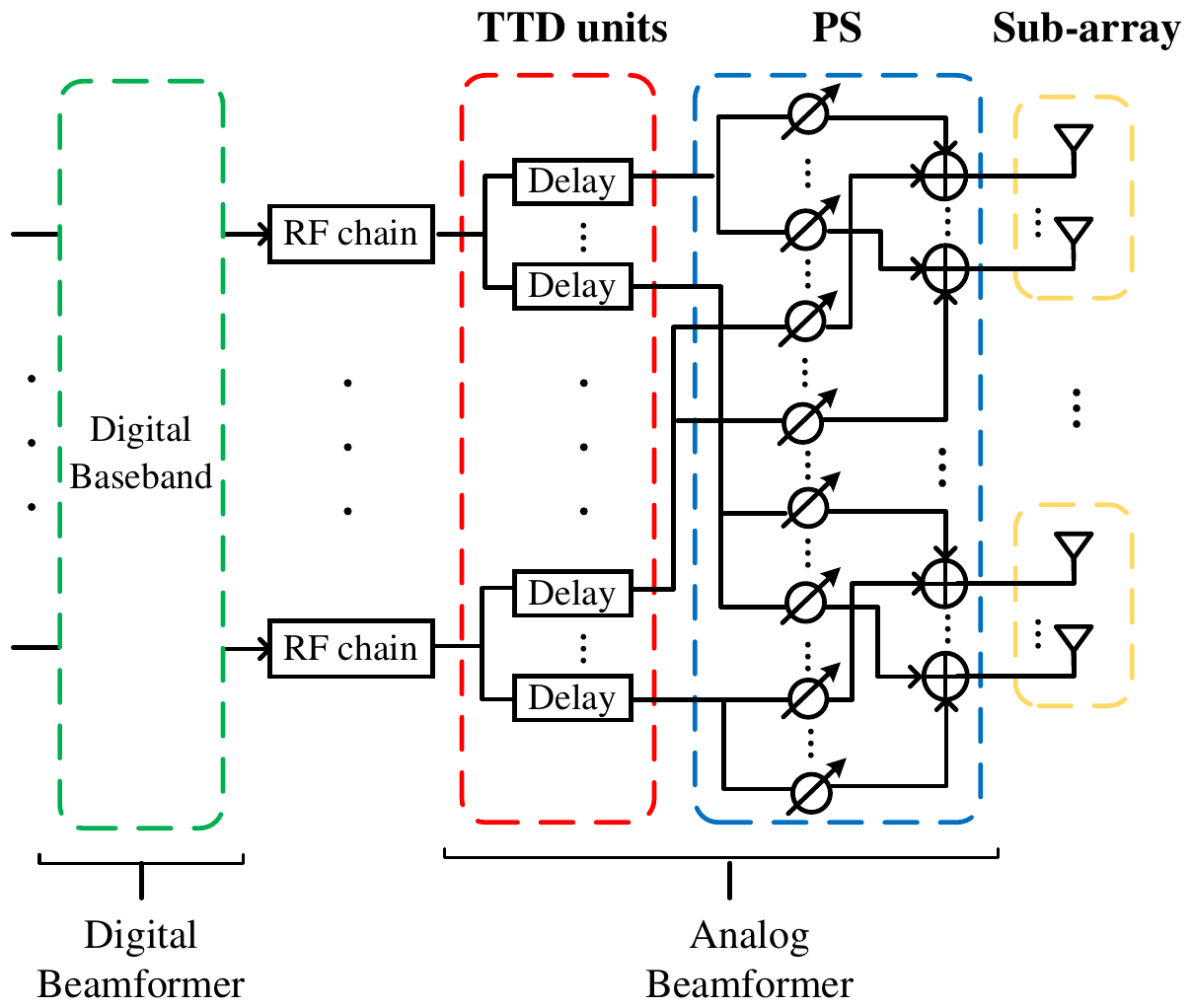}
\caption{Illustration of TTD-based hybrid beamforming structure for alleviating near-field beam squint effect in wideband UCA systems.}
\label{fig2}
\end{figure}

In particular, we consider a hybrid beamforming architecture, as depicted in Fig. \ref{fig2}. In this configuration, the $N$-elements UCA is divided into $Q$ sub-arrays with $P$ antennas per sub-array ($P = N/Q$). Furthermore, we assume that the number of RF chains is ${N_{F}}$, satisfying $K = {N_{F}} \ll N$. Each RF chain drives $Q$ TTD units, with each TTD unit connected to $P$ PSs, thereby forming a TTD-PS-based analog beamformer structure. Through manipulation of the delay in each antenna branch, the TTD units can effectively adjust the frequency-dependent phase shifts on wideband signals. For instance, we represent the adjustable time delay parameter as ${{\tau }}$, and its time-domain response can be expressed as $\delta \left( {t - \tau } \right)$. The frequency response of a TTD unit at $f_m$ is ${e^{ - j2\pi {f_m}\tau }}$. Then, we set ${r '} = c{\tau }$ as the adjustable distance parameter. The corresponding frequency-domain response can be converted to ${e^{ - j{k_m}{r '}}}$. Note that the channel phase $- {k_m}r_1^{(n)} =  - \frac{{2\pi {f_m}}}{c}r_1^{(n)}$ in equation \eqref{5} is linear in the frequency $f_m$, which can be realized by a time delay of $\frac{{r_1^{(n)}}}{c}$. Therefore, we can conclude that the TTD unit can effectively compensate for the frequency-dependent phase $- {k_m}r_1^{(n)}$ if ${r '} = - r_1^{(n)}$.

Based on the beamforming architecture combining PSs and TTDs, the received signal at $m$-subcarrier is given by
\begin{equation}
\label{12}
{{\bf{y}}_{m}} = {\bf{H}}_{m}^H{{\bf{W}}_1}{{\bf{W}}_{2,m}}{{\bf{D}}_m}{{\bf{x}}_m} + {{\bf{n}}_{m}},
\end{equation}
where ${{\bf{x}}_m} \in {\mathbb{C}^{{K} \times 1}}$ and ${{\bf{y}}_{m}} \in {\mathbb{C}^{K \times 1}}$ denote the transmitted signal and the received signal, respectively. The matrices ${\bf{H}}_{m} \in {\mathbb{C}^{N \times K}}$ and ${\bf{D}}_m \in {\mathbb{C}^{{N}_{F}\times {K}}}$ represent the channel from transmitter to users and the baseband digital beamformer on the ${m}$-th subcarrier, respectively. Herein, ${{\bf{W}}_1} \in {\mathbb{C}^{N \times Q{N}_{F}}}$ and ${{\bf{W}}_{2,m}} \in {\mathbb{C}^{Q{N}_{F} \times {N}_{F}}}$ denote the PS beamforming matrix and the time delay beamforming matrix implemented by TTDs, respectively. Moreover, ${{\bf{n}}_m} \sim {\cal C}{\cal N}\left( {0,{\sigma ^2}} \right)$ is the additive white Gaussian noise (AWGN) with ${{\sigma ^2}}$ representing the noise power.

Specifically, the analog beamformer combining PSs and TTDs at the $m$-th subcarrier could be further represented by\vspace{-1mm}
\begin{equation}
\label{13}
\begin{array}{l}
{{\bf{W}}_m} = {{\bf{W}}_1}{{\bf{W}}_{2,m}}\\
\hspace{+7.8mm} = \left[ {{\bf{W}}_{\rm{1}}^{{\rm{PS}}}, \cdots ,{\bf{W}}_{{N_F}}^{{\rm{PS}}}} \right] \times \left[ {\begin{array}{*{20}{c}}
{{\bf{W}}_{1,m}^{{\rm{TTD}}}}& \cdots &{\bf{0}}\\
 \vdots & \ddots & \vdots \\
{\bf{0}}& \cdots &{{\bf{W}}_{{N_F},m}^{{\rm{TTD}}}}
\end{array}} \right],
\end{array}
\end{equation}
where ${\bf{W}}_l^{{\rm{PS}}} \in {\mathbb{C}^{{N} \times Q}}$ and ${\bf{W}}_{l,m}^{{\rm{TTD}}}\in {\mathbb{C}^{{Q} \times 1}}$ represent the beamformers of PSs and TTDs associated with the $l$-th RF chain, respectively. For simplicity of description, we omit the numerical subscripts 1 and 2 for matrices ${\bf{W}}_l^{{\rm{PS}}}$ and ${\bf{W}}_{l,m}^{{\rm{TTD}}}$. Moreover, ${\bf{W}}_l^{{\rm{PS}}}$ can \hspace{-0.5mm}be written \hspace{-0.4mm}as a block diagonal matrix, i.e.,\vspace{-1mm}
\begin{align}
\label{14}
{\bf{W}}_l^{{\rm{PS}}} = {\rm{blkdiag}}\left[ {{{\left( {{\bf{w}}_l^{{\rm{PS}}}} \right)}_1}, \cdots ,{{\left( {{\bf{w}}_l^{{\rm{PS}}}} \right)}_Q}} \right],
\end{align}
where ${\left( {{\bf{w}}_l^{{\rm{PS}}}} \right)_q}\in {\mathbb{C}^{{P} \times 1}}$ denotes the analog beamforming of PSs attached to the $q$-th TTD unit. Additionally, the TTD-based analog beamformer is given by
\begin{align}
\label{15}
{{\bf{W}}_{l,m}^{{\rm{TTD}}}} = {\left[ {{e^{ - j2\pi {f_m}{\tau _{l,1}}}}, \cdots ,{e^{ - j2\pi {f_m}{\tau _{l,Q}}}}} \right]^T},
\end{align}
where ${\tau _{l,q}}, \hspace{0.4mm}q = 1, \cdots ,Q$ represents the time delay values of the $q$-th TTD unit connected to the $l$-th RF chain. Then, corresponding to the $l$-th RF chain, the analog beamformer is represented as ${{\bf{W}}_{l,m}} = {\bf{W}}_l^{{\rm{PS}}}{\bf{W}}_{l,m}^{{\rm{TTD}}}$, where ${{\bf{W}}_{l,m}}$ denotes the $l$-th column of ${{\bf{W}}_m}$.

Given that PSs are frequency-independent, it becomes challenging to fulfill the requirements for frequency-dependent phase shifts in wideband UCA systems. To address this, we first design the PSs to ensure that the beams align with the desired user directions at the center frequency within each sub-array. Subsequently, TTD units are designed to compensate for frequency-dependent residual components between different sub-arrays, thereby effectively mitigating the beam squint effect. In particular, the $q$-th element of ${\bf{w}}_l^{{\rm{PS}}} $ is expressed as\vspace{-1mm}
\begin{equation}
\label{16}
{\left( {{\bf{w}}_l^{{\rm{PS}}}} \right)_q} = {{\bf{a}}_c}{({r_l},{\phi _l})_q},
\end{equation}
where ${{\bf{a}}_c}{({r_l},{\phi _l})_q}\in {\mathbb{C}^{{P} \times 1}}$ is the $q$-th sub-vector of ${{\bf{a}}_c}({r_l},{\phi _l})$ in equation \eqref{5}. Following this, we provide the expression for the $n$-th element of the product of the near-field beamforming vector ${{\bf{a}}_m}({r_l},{\phi _l})$ and the steering vector ${{\bf{a}}_c}({r_l},{\phi _l})$, which can be represented by
\begin{align}
\label{17}
\hspace{-2mm}{\nu _n} = {e^{ - j\frac{{2\pi }}{c}\left( {{f_c} - {f_m}} \right)\left[ {{r_l} - R\cos \left( {{\phi _l} - \frac{{2\pi n}}{N}} \right) + \frac{{{R^2}}}{{2{r_l}}}{{\sin }^2}\left( {{\phi _l} - \frac{{2\pi n}}{N}} \right)} \right]}},
\end{align}
where $n = 0, \cdots ,N-1$. Subsequently, the corresponding $n$-th element value of ${\bf{W}}_{l,m}^{{\rm{TTD}}}$ is given by
\begin{equation}
\label{18}
{\left( {{\bf{W}}_{l,m}^{{\rm{TTD}}}} \right)_n} = {e^{j\frac{{2\pi }}{c}\left( {{f_c} - {f_m}} \right)\left[ {{r_l} - R\cos \left( {{\phi _l} - \frac{{2\pi n}}{N}} \right) + \frac{{{R^2}}}{{2{r_l}}}{{\sin }^2}\left( {{\phi _l} - \frac{{2\pi n}}{N}} \right)} \right]}}.\vspace{-1mm}
\end{equation}

Recalling that there are $Q$ TTDs, with each TTD unit is connected to a sub-array. Consequently, the $q$-th element of ${\bf{W}}_{l,m}^{{\rm{TTD}}}$ for connecting to $q$-th TTDs is unveiled in the subsequent lemma.\vspace{-1mm}
\begin{lemma} \label{lemma_3}
    \emph{The optimal design for the frequency-dependent ${\bf{W}}_{l,m}^{{\rm{TTD}}}$ implemented by the TTD unit, aiming to maximize the beamforming gain, is formulated as}\vspace{-7mm}
\end{lemma}
\begin{equation}
\label{19}
{\left( {{\bf{W}}_{l,m}^{{\rm{TTD}}}} \right)_q}= {e^{j\left( {{k_c} - {k_m}} \right)\left[ {{r_l} - R\cos \left( {{\phi _l} - {\theta _q}} \right) + \frac{{{R^2}}}{{2{r_l}}}{{\sin }^2}\left( {{\phi _l} - {\theta _q}} \right)} \right]}},\vspace{-1mm}
\end{equation}
where ${\theta _q}$ represents the phase shift of the $q$-th sub-array, which is expressed as ${\theta _q} = \frac{{\left( {P - 1} \right)\pi }}{N} + \frac{{2\pi q}}{Q}$. Furthermore, the beamforming gain can be written as\vspace{-2mm}
\begin{equation}
\label{20}
{G_m}\left({{\bf{W}}_{l,m}^{{\rm{TTD}}},{r_l},{\phi _l}} \right) \approx \frac{1}{P}\sum\limits_{j = 0}^{P - 1} {{J_0}\left( {{R_j}} \right)},\vspace{-1mm}
\end{equation}
where ${{R_j} = \sqrt 2 \left( {{k_c} - {k_m}} \right)\!R\!\left( {1 \!\!-\!\! \frac{R}{{4r_l}}} \right)\!\!\sqrt {1 \!-\! \cos \!\left( {\frac{{2\pi j }}{N} \!-\! \frac{\vartheta  }{Q}} \right)} }$.
\vspace{+1mm}
\begin{proof}
The detailed mathematical proof of \textbf{Lemma \ref{lemma_3}} can be found in Appendix \ref{proofC}.
\end{proof}

We further decompose \eqref{19} into its frequency-independent component and frequency-dependent component, denoted as
\begin{equation}
\begin{array}{l}
\label{21}
\hspace{-2.3mm}\angle \!{\left( \!{{\bf{W}}_{l,m}^{{\rm{TTD}}}}\! \right)_q} \!\!=\! {k_c}\left[ {{r_l} - R\cos \left( {{\phi _l} - {\theta _q}} \right) + \frac{{{R^2}}}{{2{r_l}}}{{\sin }^2}\left( {{\phi _l} - {\theta _q}} \right)} \right]\\
\hspace{+17.7mm} - {k_m}\!\!\left[ {{r_l} - R\cos \left( {{\phi _l} \!-\! {\theta _q}} \right) + \frac{{{R^2}}}{{2{r_l}}}{{\sin }^2}\left( {{\phi _l} - {\theta _q}} \right)} \right].
\end{array}
\end{equation}
Note that the first term in \eqref{21} can be implemented using PSs, while the second term can be realized using TTDs. Therefore, we can formulate the necessary time delay for the $q$-th TTD unit corresponding to the $l$-th channel component as \vspace{-1mm}
\begin{equation}
\label{22}
\hspace{-0.5mm}- 2\pi {f_m}{\tau _{l,q}} \!=\!  - \frac{{2\pi {f_m}}}{c}\!\!\left[ {{r_l} \!-\! R\cos \left( {{\phi _l} \!-\! {\theta _q}} \right) \!+\! \frac{{{R^2}}}{{2{r_l}}}{{\sin }^2}\left( {{\phi _l} \!-\! {\theta _q}} \right)} \!\right].
\end{equation}
Then, the time delay of the TTD-based analog beamformer can be easily deduced as\vspace{-3mm}
\begin{algorithm}[t]
        \caption{Analytical beamforming algorithm for wideband UCA systems.}\label{algorithm1}
        \begin{algorithmic}[1]
               \State {\bf{Input:}} channel matrix ${{\bf{H}}_m}$, the number of TTD units $Q$, RF chains ${N_{F}}$, frequency $f_m$, distance ${r_{l}}$, and angles ${{\phi _l}}$;
               \State~ {\bf{For}} $l = 1$ to $N_{F}$ {\bf{do}}

               \State \hspace{4.7mm} Formulate the beamsteering vector ${{{\bf{a}}_c}({r_l},{\phi _l})}$ via \eqref{5};

               \State \hspace{5mm} Obtain the TTD-based analog beamformer ${\bf{W}}_{l,m}^{{\rm{TTD}}}$ for

               \hspace{-2.2mm} the $l$-th RF chain according to \eqref{23} and \eqref{15};

               \State \hspace{4.9mm} Obtain the PS-based analog beamformer ${\bf{W}}_l^{{\rm{PS}}}$ for the

               \hspace{-1.9mm} $l$-th RF chain according to equation \eqref{14} and \eqref{24};

               \State~ {\bf{End For}}

               \State \hspace{+1.2mm} Determine the analog beamformer combining PSs and

               \hspace{-5.9mm} TTDs ${{\bf{W}}_m} $ for the m-th subcarrier via equation \eqref{13};

               \State \hspace{+1.5mm} Construct the equivalent channel ${\bf{\hat H}}_m = {\bf{H}}_{m}^H{{\bf{W}}_1}{{\bf{W}}_{2,m}}$;

               \State \hspace{+1.4mm} Based on \cite{35}, we can determine the optimal digital\vspace{1mm}

               \hspace{-5.3mm} beamformer ${{\bf{D}}_m}\!\! = \!\!\frac{{{{\left( {{{\bf{I}}} + \sum\limits_{i = 1}^K {\frac{P_t}{{K{\sigma ^2}}}} {{{\bf{\hat h}}}_{m,i}}^H{\bf{\hat h}}_{m,i}} \right)}^{ - 1}}{{{\bf{\hat h}}}_{m,k}^H}}}{{\left\| {{{\left( {{{\bf{I}}} + \sum\limits_{i = 1}^K {\frac{P_t}{{K{\sigma ^2}}}} {{{\bf{\hat h}}}_{m,i}^H}{\bf{\hat h}}_{m,i}} \right)}^{ - 1}}{{{\bf{\hat h}}}_{m,k}^H}} \right\|}}$, where

                \hspace{-4.1mm}${{{{\bf{\hat h}}}_{m,k}}}$ is the $k$-th row of the equivalent channel ${\bf{\hat H}}_m$;\vspace{1mm}

               \State {\bf{Output:}} ${{\bf{W}}_1}$, ${{\bf{W}}_{2,m}}$ and ${{\bf{D}}_m}$.
        \end{algorithmic}
\end{algorithm}
\begin{equation}
\label{23}
{\tau _{l,q}} = \frac{1}{c}\left[ {{r_l} - R\cos \left( {{\phi _l} - {\theta _q}} \right) + \frac{{{R^2}}}{{2{r_l}}}{{\sin }^2}\left( {{\phi _l} - {\theta _q}} \right)} \right].\vspace{-1.5mm}
\end{equation}

Furthermore, by combining equation \eqref{16} with the frequency-independent component in equation \eqref{21}, we can reformulate the beamformer of the PSs associated with the $l$-th RF chain as\vspace{-1.5mm}
\begin{equation}
\label{24}
{\left( {{\bf{w}}_l^{{\rm{PS}}}} \right)_q} = {{\bf{a}}_c}{({r_l},{\phi _l})_q}{e^{j{k_c}\left[ {{r_l} - R\cos \left( {{\phi _l} - {\theta _q}} \right) + \frac{{{R^2}}}{{2{r_l}}}{{\sin }^2}\left( {{\phi _l} - {\theta _q}} \right)} \right]}}.\vspace{-1mm}
\end{equation}

The proposed TTD-based analytical beamforming algorithm is summarized in \textbf{Algorithm \ref{algorithm1}}. It is worth highlighting that obtaining all channel components, especially $\phi_l$ and $r_l$, is considered as a prerequisite for the proposed analytical algorithm, which is commonly regarded as a fundamental condition in prior research \cite{9, 15}. Specifically, in near-field mmWave communications, the channel state information (CSI) can be easily obtained by deep learning-based approach \cite{33} or compressive sensing approach \cite{34}.

\subsection{Design the Number of TTD Units}
Based on the aforementioned analog beamforming analysis, we further investigate how many TTD units are required to effectively alleviate the beam squint effect across all subcarriers in this subsection.

Utilizing the Taylor series expansion $\cos (x)=1-\frac{1}{2} x^2+\mathcal{O}\left(x^2\right)$, $R_j$ in \eqref{20} can be further reformulated as\vspace{-1.5mm}
\begin{equation}
\begin{array}{l}
\label{25}
{R_j} = \sqrt 2 \left( {{k_c} - {k_m}} \right)R\left( {1 - \frac{R}{{4r_l}}} \right)\sqrt {1 - \cos {\kappa _j}} \\
\hspace{+5.1mm} \mathop  \approx \left( {{k_c} - {k_m}} \right)R\left( {1 - \frac{R}{{4r_l}}} \right){\kappa _j},\vspace{-2mm}
\end{array}
\end{equation}
where ${\kappa _j} = {\frac{{2\pi j}}{N} - \frac{\vartheta }{Q}} $ and $j = 0, \cdots ,P - 1$. Subsequently, the beamforming gain ${G_m}\left( {{\bf{W}}_{l,m}^{{\rm{TTD}}},{r_l},{\phi _l}} \right)$ could be further simplified as\vspace{-3mm}
\begin{align}
\label{26}
\hspace{-0.2mm}{G_m}\!\!\left( \!{{\bf{W}}_{l,m}^{{\rm{TTD}}},{r_l},{\phi _l}}\! \right) & \!= \! \frac{1}{P}\!\!\sum\limits_{j = 0}^{P - 1} \!\!{{J_0}\! \left( { \left( {{k_c} - {k_m}} \right)\left({R - \frac{R^2}{{4r_l}}} \right){\kappa _j}} \right)} \nonumber \\
&\! \mathop  \approx \limits^{\left( a \right)} \! \frac{N}{{2\pi P}}\!\!\!  \int_{ - \pi /Q}^{\pi /Q}\!\!\! {{J_0}} \! \left(\!\! { \left( {{k_c} \!-\!{k_m}} \right)\!\left(\!\! {R - \frac{R^2}{{4r_l}}} \right)\!\kappa } \!\right)\!\!{\rm{d}}\kappa \nonumber\\
&\hspace{-0.5mm} \mathop  = \limits^{\left( b \right)}\!\! \frac{N}{{\pi P}}\!\!\!\int_0^{\pi /Q}\!\! {{J_0}}\!  \!\left( \!{ \left( {{k_c} - {k_m}} \right)\left( {\!R - \frac{R^2}{{4r_l}}} \right)\!\kappa } \right)\!\!{\rm{d}}\kappa,
\end{align}
where ${\left( a \right)}$ is derived by replacing summation with integral over $\kappa $ and ${\left( b \right)}$ stems from the parity property of the Bessel function ${J_0}\left(  \cdot  \right)$ \cite{32}. Substituting $\kappa ' = \left( {{k_c} - {k_m}} \right)R\left( {1 - \frac{R}{{4r_l}}} \right)\kappa $ and $\varepsilon \!=\! \frac{\pi }{Q} \cdot \left( {{k_c} \!-\! {k_m}} \right)R\left( {1 \!-\! \frac{R}{{4r_l}}} \right)$, equation \eqref{26} can be derived as \vspace{-4mm}
\begin{align}
\label{27}
{G_m}\left( {{\bf{W}}_{l,m}^{{\rm{TTD}}},{r_l},{\phi _l}} \right)  &\approx \frac{1}{\varepsilon }\int_0^\varepsilon {{J_0}} \left( {\kappa '} \right){\rm{d}}\kappa ' \nonumber\\
 & \mathop  = \limits^{\left( c \right)} {}_1{F_2}\left( {\frac{1}{2};1,\frac{3}{2}; - \frac{{{\varepsilon ^2}}}{4}} \right),
\end{align}
where ${}_1{F_2}$ represents the generalized hypergeometric function. Formula (c) is derived based on the integral property of Bessel functions. It is evident that the function ${}_1{F_2}$ is a one-variable function with respect to $\varepsilon$. In particular, as the variable $\varepsilon$ increases, the function ${}_1{F_2}$ exhibits the oscillatory downward trend. As a result, we can deduce that a greater quantity of TTD units is necessary for achieving a more pronounced reduction in beamforming loss.

However, a substantial number of TTD units leads to increased hardware complexity and higher power consumption. To address this, we estimate the required number of TTD units by setting a predetermined threshold $\Delta$, with the objective of maintaining the beamforming loss below that threshold. Recalling the definition of $\varepsilon$ in equation \eqref{27}, and based on the condition $\left| {{f_c} - {f_m}} \right| \le \frac{B}{2}$ holds for any subcarrier $f_m$, we can easily derive the necessary number of TTD units as \vspace{-1mm}
\begin{equation}
\label{28}
Q \ge \frac{{ {\pi}^2B}}{{c{f^{ - 1}}\left( {1 - \Delta} \right)}}R\left( {1 - \frac{R}{{4{r_l}}}} \right),
\end{equation}
where ${f^{ - 1}}\left( {1 - \Delta} \right)$ can be obtained as follows \vspace{-1mm}
\begin{equation}
\label{29}
{f^{ - 1}}\left( {1 - \Delta} \right) \buildrel \Delta \over = \mathop {\arg \min }\limits_\varepsilon  \left\{ {{}_1{F_2}\left( {\frac{1}{2};1,\frac{3}{2}; - \frac{{{\varepsilon^2}}}{4}} \right) = 1 - \Delta } \right\}.
\end{equation}
It can be observed that as the distance between the user and transmitter increases, the required number of TTD units also increases. This is because the greater distance between the transmitter and receiver leads to enhanced multipath effects and beam squint effects, which may necessitate a greater number of TTD units for compensation and optimization.

\section{J\footnotesize{OINT} \normalsize{D}\footnotesize{ELAY} AND \normalsize{P}\footnotesize{HASE} \normalsize{B}\footnotesize{EAMFORMING} UNDER \normalsize{TTD} \normalsize{C}\footnotesize{ONSTRAINTS}}\label{sectionV}
The analysis in Section \ref{sectionIV} primarily concentrates on the design of the analog beamformer generated by TTD units while keeping the PSs beamformer fixed. This simplifies the analog beamforming design as it decouples TTD and PS beamformers. Additionally, it is important to highlight that in Section \ref{sectionIV}, we assume that the time delay values generated by TTD units are unconstrained and can increase infinitely without any constraints. However, achieving this becomes challenging since the assumption of unbounded time delay cannot be realized in practice. The range of time delay values is strictly limited. As a result, in contrast to the design with fixed PSs and optimized TTDs, performing a joint optimization of both the TTD and PS beamformers under the specific time delay constraints of TTD units is necessary. Through the joint control of the delay and phase, the beams across the entire bandwidth are precisely focused on the desired user locations, thereby effectively mitigating the beam squint effect.

To be specific, we adopt a two-step joint-optimization method for the analog beamformer design. Firstly, we characterize the optimal unconstrained analog beamformer that effectively compensates for the near-field beam squint effect in wideband UCA systems. Subsequently, the problem can be formulated as a joint TTD and PS optimization problem aimed at minimizing the discrepancy between the product of PS and TTD beamformers and the optimal unconstrained analog beamformer. The basic principle of this method is to closely approximate the optimal unconstrained analog beamformer, thereby maximizing the performance of near-field wideband communication systems.

\subsection{Optimal Unconstrained Analog Beamformer}\label{sectionVA}
We first determine the optimal unconstrained analog beamformer that maximizes the array gain of each subcarrier. The primary purpose of determining the optimal analog beamformer aims to present a reference design for the TTD and PS joint beamforming scheme proposed in the subsequent section.

Specifically, we define ${{{\bf{\tilde W}}}_m} \in {\mathbb{C}^{{N} \times {N}_{F}}}$ as the unconstrained analog beamformer, where ${{{\bf{\tilde w}}}_{l,m}}\in {\mathbb{C}^{{N} \times 1}}$ is the $l$-th column of \hspace{-0.3mm}${{{\bf{\tilde W}}}_m}$. \hspace{-0.7mm}Then, the \hspace{-0.3mm}array \hspace{-0.3mm}gain \hspace{-0.3mm}achieved \hspace{-0.5mm}by \hspace{-0.5mm}${{{\bf{\tilde w}}}_{l,m}}$ \hspace{-0.5mm}can \hspace{-0.5mm}be \hspace{-0.5mm}written \hspace{-0.5mm}as
\begin{equation}
\label{30}
{G_m}\left( {{{\bf{a}}_m}({r_l},{\phi _l}),{{{\bf{\tilde w}}}_{l,m}}} \right) = \left| {{\bf{a}}_m^H({r_l},{\phi _l}){{{\bf{\tilde w}}}_{l,m}}} \right|,
\end{equation}
where ${{\bf{a}}_m}({r_l},{\phi _l})$ represents the array response vector of the $l$-th path at the $m$-th subcarrier. Utilizing the Cauchy inequality, we obtain
\begin{equation}
\label{31}
{G_m}\left( {{{\bf{a}}_m}({r_l},{\phi _l}),{{{\bf{\tilde w}}}_{l,m}}} \right) \le {\left\| {{{\bf{a}}_m}({r_l},{\phi _l})} \right\|_2}{\left\| {{{{\bf{\tilde w}}}_{l,m}}} \right\|_2} = 1.
\end{equation}
It is noteworthy to point out that the equality holds if and only if ${{{\bf{\tilde w}}}_{l,m}} = {{\bf{a}}_m}({r_l},{\phi _l})$, resulting in ${\bf{\tilde W}}_m^ *  = \left[ {{{\bf{a}}_m}({r_1},{\phi _1}), \cdots ,{{\bf{a}}_m}({r_l},{\phi _l})} \right]$. In particular, the entry in the $n$-th row and $l$-th column of the optimal unconstrained analog beamforming matrix ${\bf{\tilde W}}_m^*$ is given by
\begin{equation}
\label{32}
\tilde W_m^ * \left( {n,l} \right)= \frac{1}{{\sqrt N }}{e^{-j{k_m}\left( {{r_l} + \varsigma _{{r_l},{\phi _l}}^{(n)}} \right)}}.
\end{equation}
Note that achieving the optimal analog beamformer ${\bf{\tilde W}}_m^ * $ in \eqref{32} necessitates equipping each transmit antenna with a TTD unit, i.e., $Q = N$ and $P = 1$. However, this design is highly impractical for XL-MIMO system as it necessitates a substantial number of TTD units, leading to increased hardware complexity and higher power consumption. To address this issue, we propose a method that jointly optimizes the PS and TTD beamformers to approximate the optimal unconstrained analog beamformer.

\subsection{Joint Optimization of the TTD and PS beamformers}
Based on the aforementioned analysis, it is desired to find ${{\bf{W}}_1}$ and ${{\bf{W}}_{2,m}}$ such that ${{\bf{W}}_1}{{\bf{W}}_{2,m}} = {\bf{\tilde W}}_m^ *$. However, solving $M$-coupled matrix equations through mathematical derivation is quite challenging. Consequently, we intend to minimize the sum of mean square errors between the optimal unconstrained analog beamformer and the analog counterparts across all subcarriers. The optimization problem is given by\vspace{-1.5mm}
\begin{subequations}
\label{33}
\begin{align}
\label{33a}
\vspace{-4mm}
\min _{{\bf{W}}_l^{{\rm{PS}}},{{\bf{W}}_{l,m}^{{\rm{TTD}}}}}\!\!\!\!\! &\ \ \ \Upsilon  \triangleq \sum_{m \in M}\left\|{\bf{\tilde W}}_m^ *-{\bf{W}}_l^{{\rm{PS}}}{{\bf{W}}_{l,m}^{{\rm{TTD}}}}\right\|^2 \\
\label{33b}
\text { s.t. }\ &\ \ \left|\left[{\bf{W}}_l^{{\rm{PS}}}\right]_n\right|=1, \forall n \in \mathcal{N} \triangleq\{1,\cdots , N\}, \\
\label{33c}
& \ \ \ {\bf{W}}_{l,m}^{{\rm{TTD}}} = F\left( {{f_m},{{\bf{t}}_1}, \cdots ,{{\bf{t}}_{{N_F}}}} \right),\\
\label{33d}
& \ \ \ {{\bf{t}}_l} \in {\mathcal{T}_l},\forall l \in \mathcal{L}\buildrel \Delta \over = \left\{ {1,\cdots ,{N_F}} \right\},\\
\label{33e}
& \ \ \ 0 \leq \tau_{l,q} \leq \tau_{\max }, \forall q \in \mathcal{Q} \triangleq\{1, \cdots , Q\},
\end{align}
\end{subequations}
where constraint \eqref{33b} represents the unit-modulus constraint of the PS-based analog beamformer. Constraints \eqref{33c}-\eqref{33e} indicate the hardware limitations associated with TTD-based analog beamformers. Herein, $\mathbf{t}_l=\left[\tau_{l, 1}, \cdots, \tau_{l, Q}\right]^T \in \mathbb{R}^{Q \times 1}$ and ${\mathcal{T}_l}$ denotes the feasible set of the time delay vector for the TTD units attached to the $l$-th RF chain. Moreover, $\tau_{\max }$ signifies the maximum time delay value of the TTD unit. Then, we further transform the objective function of \eqref{33} into
\begin{equation}
\label{34}
\begin{array}{l}
\hspace{-1.5mm}{\left\| {{\bf{\tilde W}}_m^ * - {\bf{W}}_l^{{\rm{PS}}}{{\bf{t}}_l}} \right\|^2} = \sum\limits_{q = 0}^{Q - 1} {{{\left\| { {\bf{\tilde w}}_{l,m}^* - {\bf{w}}_l^{{\rm{PS}}}\exp \left( { - j2\pi {f_m}{\tau _{l,q}}} \right)} \right\|}^2}} \\
\hspace{+3.5mm} = {V_{l,m}} + \sum\limits_{q = 0}^{Q - 1}  -  2{\mathop{\rm Re}\nolimits} \left\{ {{{\left( {{\bf{\tilde w}}_{l,m}^*} \right)}^H}{\bf{w}}_l^{{\rm{PS}}}\exp \left( { - j2\pi {f_m}{\tau _{l,q}}} \right)} \right\},
\end{array}
\end{equation}
where ${V_{l,m}} \!=\! \sum\limits_{q = 0}^{Q - 1} {\left( \!{{{\left( { {\bf{\tilde w}}_{l,m}^*} \right)}^H} {\bf{\tilde w}}_{l,m}^* \!+\! {{\left( {{\bf{w}}_l^{{\rm{PS}}}} \right)}^H}{\bf{w}}_l^{{\rm{PS}}}} \right)}$ and ${{{{\bf{\tilde w}}}_{l,m}}^ *}$ is the $l$-th column of ${\bf{\tilde W}}_m^*$. In particular, ${{{{\bf{\tilde w}}}_{l,m}}^*}$ can be determined based on the analysis in Section \ref{sectionVA}. Furthermore, considering the unit-modulus constraint of ${{\bf{w}}_l^{{\rm{PS}}}}$, it is evident that ${\left( {{\bf{w}}_l^{{\rm{PS}}}} \right)^H}{\bf{w}}_l^{{\rm{PS}}} = P$, which remains constant. As a result, the optimization variable is only associated with the second term, which is the primary focus of our analysis. Subsequently, we can reformulate \eqref{33} as the following equivalent problem \vspace{-0.5mm}
\begin{subequations}
\begin{align}
\label{35a}
\max _{{\bf{w}}_l^{{\rm{PS}}}, \tau_{l,q}} & \ \sum\limits_{m \in M} {{\mathop{\rm Re}\nolimits} \left\{ {{{\left( {{\bf{\tilde w}}_{l,m}^*} \right)}^H}{\bf{w}}_l^{{\rm{PS}}}\exp \left( { - j2\pi {f_m}{\tau _{l,q}}} \right)} \right\}}  \\
\text { s.t. }\ & \  \  \eqref{33b},\eqref{33c},\eqref{33d},\eqref{33e}.
\end{align}
\end{subequations}
It can be observed that the PSs and TTDs analog beamforming matrices are coupled in the objective function \eqref{35a}, making it challenging to optimize PSs and TTDs simultaneously. Consequently, to overcome this, we adopt an alternating optimization strategy, optimizing ${\bf{w}}_l^{{\rm{PS}}}$ and $\tau_{l,q}$ individually.
\begin{algorithm}[t]
        \caption{Joint-optimization beamforming algorithm for wideband UCA systems} \label{algorithm2}\vspace{-0.2mm}
        \begin{algorithmic}[1]
               \State {\bf{Input:}} a predefined threshold $\Upsilon $ and channel matrix ${{\bf{H}}_m}$

               \State {\bf{Initialize:}} feasible optimization variables ${{{\bf{\tilde W}}}_{m}}$, ${\bf{W}}_l^{{\rm{PS}}}$, and

               \hspace{7mm} ${{\bf{W}}_{l,m}^{{\rm{TTD}}}}$;

               \State \hspace{1.2mm} {\bf{For}} $l = 1$ to $N_{F}$ {\bf{do}}\vspace{-0.5mm}

               \State \hspace{3.6mm} {\bf{Repeat}}\vspace{-0.5mm}

               \State \hspace{6.7mm} Update the PS-based beamformer ${\bf{W}}_l^{{\rm{PS}}}$ via \eqref{37};

               \State \hspace{7.1mm} Update the TTD-based beamformer ${{\bf{W}}_{l,m}^{{\rm{TTD}}}}$ via \eqref{39};

               \State \hspace{3.8mm} {\bf Until} the objective function value of (33{\rm{a}}) falls below

               \State \hspace{4.2mm} a predefined threshold $\Upsilon $.
               \State \hspace{1.2mm} {\bf{End For}}\vspace{-0.5mm}

               \State \hspace{1.2mm} Determine the analog beamformer combining PSs and

               \hspace{-5.9mm} TTDs ${{\bf{W}}_m} $ for the m-th subcarrier via equation \eqref{13};

               \State\hspace{1.2mm} Construct the equivalent channel ${\bf{\hat H}}_m = {\bf{H}}_{m}^H{{\bf{W}}_1}{{\bf{W}}_{2,m}}$;

               \State \hspace{+1.3mm} Based on \cite{35}, we can determine the optimal digital\vspace{-0.5mm}

               \hspace{-5.5mm} beamformer ${{\bf{D}}_m} = \frac{{{{\left( {{{\bf{I}}} + \sum\limits_{i = 1}^K {\frac{P_t}{{K{\sigma ^2}}}} {{{\bf{\hat h}}}_{m,i}}^H{\bf{\hat h}}_{m,i}} \right)}^{ - 1}}{{{\bf{\hat h}}}_{m,k}^H}}}{{\left\| {{{\left( {{{\bf{I}}} + \sum\limits_{i = 1}^K {\frac{P_t}{{K{\sigma ^2}}}} {{{\bf{\hat h}}}_{m,i}^H}{\bf{\hat h}}_{m,i}} \right)}^{ - 1}}{{{\bf{\hat h}}}_{m,k}^H}} \right\|}}$;

               \State {\bf{Output:}} ${{\bf{W}}_1}$, ${{\bf{W}}_{2,m}}$ and ${{\bf{D}}_m}$.\vspace{-1mm}
        \end{algorithmic}
\end{algorithm}

\subsubsection{Subproblem of Optimizing PSs}
Given the TTDs, the subproblem of optimizing PSs ${\bf{w}}_l^{{\rm{PS}}}$ can be constructed as
\begin{subequations}
\begin{align}
\label{36}
\max _{{\bf{w}}_l^{{\rm{PS}}} } & \ \sum\limits_{m \in M} {{\mathop{\rm Re}\nolimits} \left\{ {{{\left( {{\bf{\tilde w}}_{l,m}^*} \right)}^H}{\bf{w}}_l^{{\rm{PS}}}\exp \left( { - j2\pi {f_m}{\tau _{l,q}}} \right)} \right\}}\\
\text { s.t. } & \ \ \eqref{33b}.
\end{align}
\end{subequations}
The optimal PS-based beamformer ${\bf{w}}_l^{{\rm{PS}}}$ for the aforementioned problem can be easily derived as \vspace{-1mm}
\begin{equation}
\label{37}
{\left( {{\bf{w}}_l^{{\rm{PS}}}} \right)^ * }  = {e^{j\angle \left( {\sum\limits_{m \in M} {{{{\bf{\tilde w}}}_{l,m}^*}{e^{j2\pi {f_m}{\tau _{l,q}}}}} } \right)}}. \vspace{-1mm}
\end{equation}

\subsubsection{Subproblem of Optimizing TTDs}
For any given PSs, the subproblem of optimizing TTDs is described as
\begin{subequations}
\label{38}
\begin{align}
\max _{\tau_{l,q}} & \ \sum\limits_{m \in M} {{\rm{Re}}\left[ {\sum\limits_{q = 0}^{Q - 1} {\sum\limits_{p = 0}^{P - 1} {{\psi _{m,q,qP + p}}{e^{ - j2\pi {f_m}{\tau _{l,q}}}}} } } \right]}  \\
\text { s.t. } &\ \ \ {\tau _{l,q}} \in \left[ {0,{\tau _{\max }}} \right],
\end{align}
\end{subequations}
where ${\psi _{m,q,p}} = {{{\left( {\widetilde {\bf{w}}_{l,m}^*} \right)}^H}{\bf{w}}_l^{{\rm{PS}}}}$. The problem presented in \eqref{38} is a classical single-variable optimization problem within a fixed interval. Consequently, the near-optimal solution for ${{\tau _{l,q}}}$ can be effectively derived using the one-dimensional search. To be specific, let $X$ represent the number of search steps, resulting in the search set ${\cal X} = \left\{ {0,\frac{{{\tau _{\max }}}}{{X - 1}},\frac{{2{\tau _{\max }}}}{{X - 1}}, \cdots ,{\tau _{\max }}} \right\}$. Then, a near-optimal solution for ${{\tau _{l,q}}}$ can be expressed as\vspace{-2mm}
\begin{equation}
\label{39}
{\tau _{l,q}} = \mathop {\arg \max }\limits_{{\tau _{l,q}} \in {\cal X}} \sum\limits_{m \in M} {{\rm{Re}}\left[ {\sum\limits_{q = 0}^{Q - 1} {\sum\limits_{p = 0}^{P - 1} {{\psi _{m,q,qP + p}}{e^{ - j2\pi {f_m}{\tau _{l,q}}}}} } } \right]}.
\end{equation}

The complete algorithm for solving the optimization problem \eqref{33} is summarized in \textbf{Algorithm \ref{algorithm2}}. Since the objective function value $\Upsilon $ is non-increasing at each step of the alternating iteration optimization, the convergence of \textbf{Algorithm \ref{algorithm2}} can be guaranteed. Specifically, \textbf{Algorithm \ref{algorithm2}} demonstrates remarkable computational efficiency by updating the optimization variables in each step using either closed-form solutions or low-complexity linear search approach. The complexity of calculating ${{{\bf{\tilde W}}}_m}$ is ${\cal O}\left( {MN} \right)$. Additionally, the complexities of updating ${\bf{W}}_{l,m}^{{\rm{TTD}}}$, ${\bf{W}}_l^{{\rm{PS}}}$, and ${{\bf{D}}_m}$ can be calculated as ${\cal O}\left( {M{N^2}{N_F}X/Q} \right)$, ${\cal O}\left( {MN} \right)$, and ${\cal O}\left( {M\left( {N_F^3 +KN_F^2 + 2{N_F}} \right)} \right)$, respectively.
\begin{figure}[t]
\centering
\vspace{-3mm}
\subfigure[ULA]{\includegraphics[width = 2.62in]{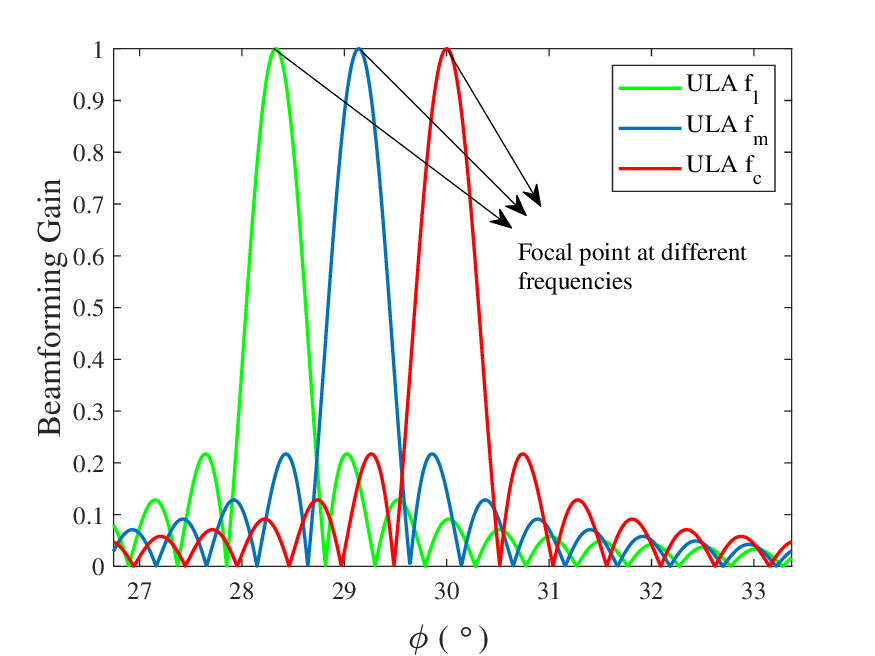}}\vspace{-3.6mm}
\subfigure[UCA]{\includegraphics[width = 2.62in]{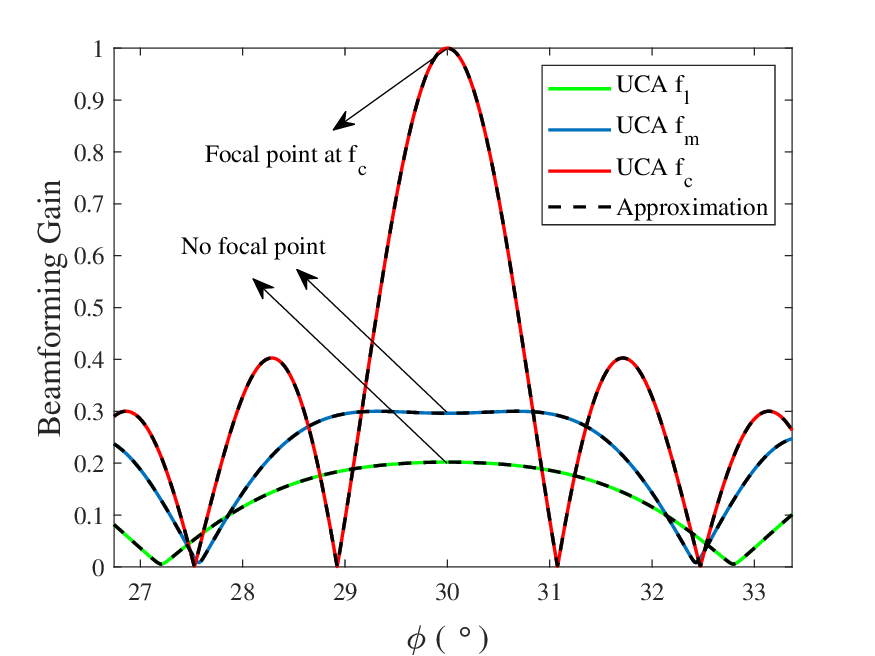}}\vspace{-1.5mm}
\caption{Illustration of beamforming gain in the angular domain. The beams are generated towards $\phi  = {30^ \circ }$ at the central frequency $f_c$ = 28 GHz. Moreover, $f_l$ = 26.5 GHz and ${f_m} = \frac{{{f_c} + {f_l}}}{2}$ denote the lowest and the middle frequency.} 
\label{fig3}
\end{figure}
\begin{figure}[t]
\centering
\vspace{-4.5mm}
\subfigure[ULA]{\includegraphics[width = 2.6in]{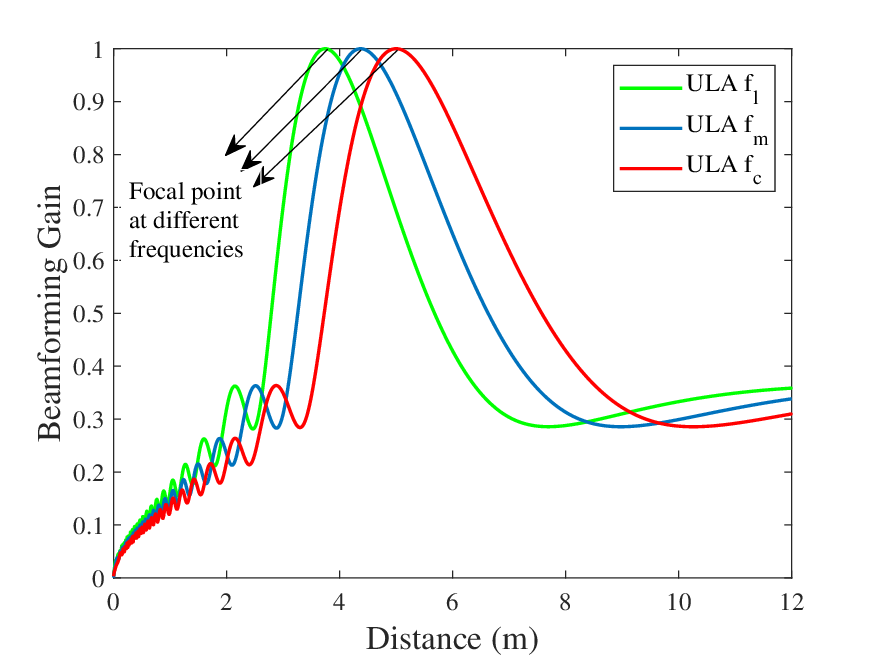}}\vspace{-3.4mm}
\subfigure[UCA]{\includegraphics[width = 2.6in]{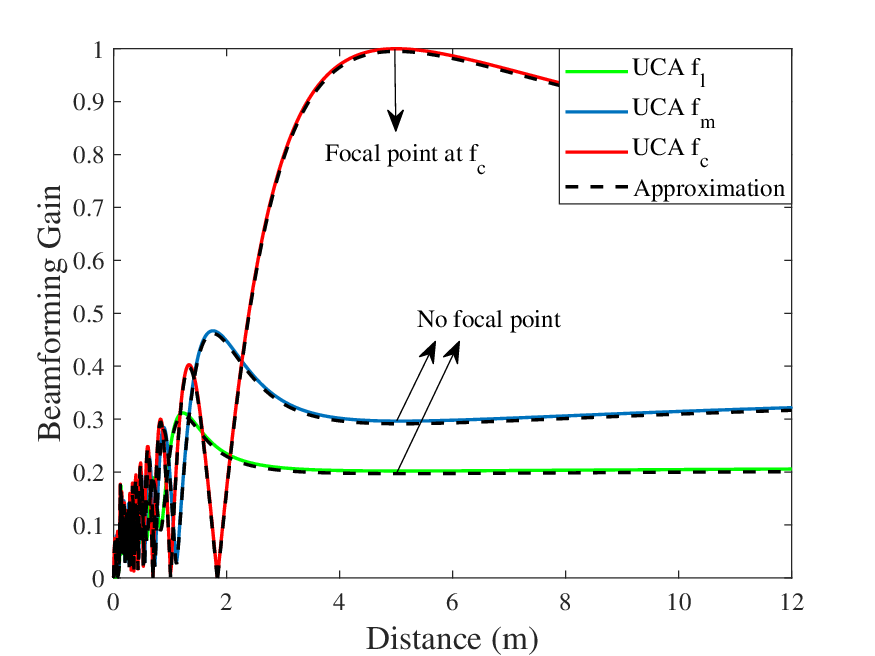}}\vspace{-1.4mm}
\caption{Illustration of beamforming gain in the distance domain.} 
\label{fig4}
\end{figure}

\section{\normalsize{N}\footnotesize{UMERICAL} \normalsize{R}\footnotesize{ESULTS}}\label{sectionVI}
In this section, numerical results are conducted to evaluate the effectiveness of the proposed analytical and joint-optimization beamforming approaches. Without special instructions, the fundamental simulation parameters in this paper are provided as follows. The central frequency, system bandwidth, and number of subcarriers are set to $f_c$ \!=\! 28 GHz, $B$ \!=\! 3 GHz, and $M$ \!=\! 10, respectively. The BS is equipped with $N$ = 256 antennas placed at a half-wavelength spacing, resulting in a Rayleigh distance of approximately ${d_{r}} = 2{R^2}/{\lambda _c} \approx 35$ m. Additionally, we assume $K$ = 4 users randomly distributed in the near-field region, with the number of RF chains equal to the number of users. Each RF chain is configured with 16 TTD units, and each TTD unit consists of 16 PSs. The maximum time delay value of each TTD unit is specified as $\tau_{\max }$ = 20 ns \cite{36}. The signal-to-noise ratio (SNR) is set to 15 dB, which is determined by the ratio of the transmit power $P_t$ to the noise power ${\sigma ^2}$. All the numerical results presented below are obtained by averaging over 200 random channel realizations.

\subsection{Analysis of Near-Field Beam Squint Effect}\label{sectionVIA}
Fig. \ref{fig3} depicts the beamforming gain for UCA and ULA in the angular domain. It is evident that the high-gain beams of the ULA exhibit a slight squint from the desired direction while maintaining the same beam pattern, as depicted in Fig. \ref{fig3}(a). On the contrary, Fig. \ref{fig3}(b) clearly demonstrates that the beam pattern of UCA experiences significant distortion at non-central frequencies, primarily due to the near-field beam squint effect. In other words, high-gain beams no longer exist in the azimuth plane, which is consistent with the analysis results derived in \textbf{Lemma \ref{lemma_1}}. Meanwhile, compared to the ULA, the beamforming pattern of the UCA exhibits a wider main lobe, facilitating more extensive spatial coverage. Furthermore, in Fig. \ref{fig3}(b), the solid lines represent the precisely calculated beamforming gain, whereas the black dashed lines represent the approximated beamforming gain using equation \eqref{8}. Obviously, the approximate results align with the accurately calculated beamforming gain. These simulation results further validate the accuracy of our previous mathematical derivations.

Fig. \ref{fig4} analyzes the beamforming gain for UCA and ULA in the distance domain. The focal point is specifically chosen at a distance of $r$ = 5 m and an angle of $\phi = {0^ \circ }$. In Fig. \ref{fig4}(a), it can be seen that the high-gain beams of the ULA show a slight deviation, while the beam pattern remains unchanged. In contrast, Fig. \ref{fig4}(b) demonstrates the significant distortion in the beam patterns of the UCA at non-central frequencies. In addition, it can also be observed that a larger distance between the beam focal point and the user results in a smaller beamforming gain. This phenomenon emphasizes the precise beamforming design in the distance domain is essential for near-field communications. Furthermore, as depicted in Fig. \ref{fig4}(b), the black dashed lines representing the beamforming gain calculated using equation \eqref{11} exhibit consistency with the precisely calculated beamforming gain, which confirms the accuracy of the approximation presented in \textbf{Lemma \ref{lemma_2}}.

\begin{figure}[t]
\centering
\vspace{-0.5mm}
\includegraphics[width=0.365\textwidth]{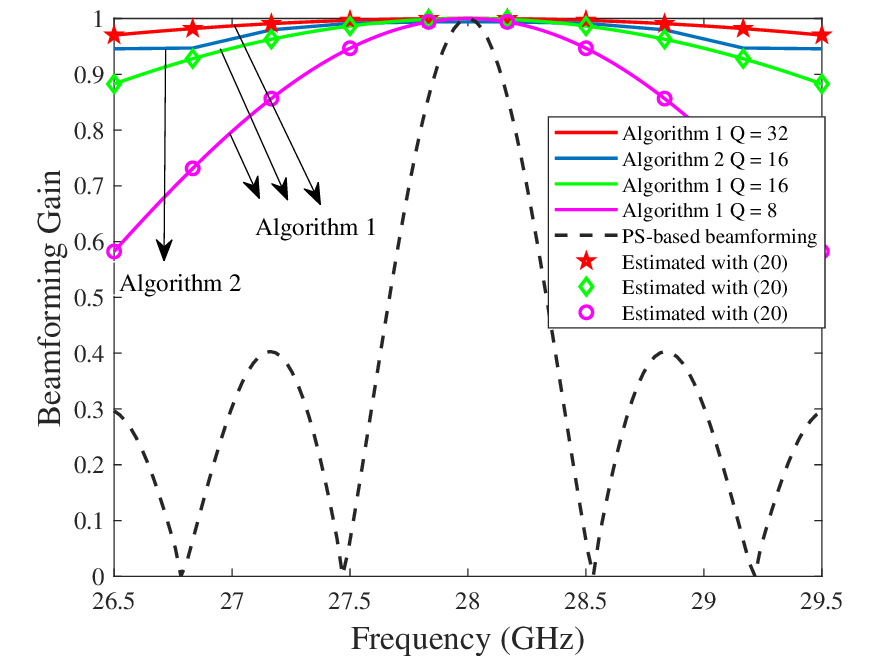}\vspace{-1.3mm}
\caption{The beamforming gain comparison between conventional PS-based beamforming scheme and TTD-based beamforming schemes.}
\label{fig5}
\end{figure}

The effectiveness of the TTD-based beamforming schemes proposed in \textbf{Algorithm \ref{algorithm1}} and \textbf{Algorithm \ref{algorithm2}} is validated in Fig. \ref{fig5}. To be specific, Fig. \ref{fig5} compares the beamforming gain between conventional PS-based beamforming, analytical beamforming, and joint-optimization beamforming. As can be seen, the analytical beamforming scheme presented in \textbf{Algorithm \ref{algorithm1}} significantly enhances the performance of beamforming gain compared to conventional PS-based beamforming scheme, especially as the number of TTD units increases. In particular, the beamforming gain has approximately reached the thresholds of 0.97, 0.89, and 0.59 over the bandwidth when the number of TTD units is 32, 16, and 8, respectively. Meanwhile, in any case, the theoretical results estimated with \textbf{Lemma \ref{lemma_3}} are consistent with the simulation results, confirming the correctness of our mathematical derivations. Additionally, the joint-optimization beamforming scheme proposed in \textbf{Algorithm \ref{algorithm2}} exhibits superior beamforming gain performance when compared to \textbf{Algorithm \ref{algorithm1}} and the PS-based beamforming scheme. This is attributed to the dynamic adjustment of phase and delay parameters during the alternating optimization of PSs and TTDs, significantly improving system performance.

\subsection{Evaluation of Spectral Efficiency}
Furthermore, we evaluate the spectral efficiency performance for multi-user systems. Specifically, the spectral efficiency of user $k$ at the $m$-th subcarrier is given by\vspace{-2mm}
\begin{equation}
\label{40}
\small
{R_{m,k}} = {\log _2}\left( {1 + \frac{{{{\left| {{\bf{h}}_{m,k}^H{{\bf{W}}_1}{{\bf{W}}_{2,m}}{{\bf{d}}_{m,k}}} \right|}^2}}}{{\sum\limits_{i = 1,i \ne k}^K {{{\left| {{\bf{h}}_{m,k}^H{{\bf{W}}_1}{{\bf{W}}_{2,m}}{{\bf{d}}_{m,k}}} \right|}^2} + {\sigma ^2}} }}} \right),
\end{equation}
where ${\bf{h}}_{m,k}$ represents the channel between user $k$ and the transmitter. Moreover, ${{{\bf{d}}_{m,k}}}$ is the $k$-th column of ${{\bf{D}}_m}$ and serves as the digital beamformer for user $k$ at $m$-th subcarrier.

\begin{figure}[t]
\centering
\vspace{-2.5mm}
\includegraphics[width=0.385\textwidth]{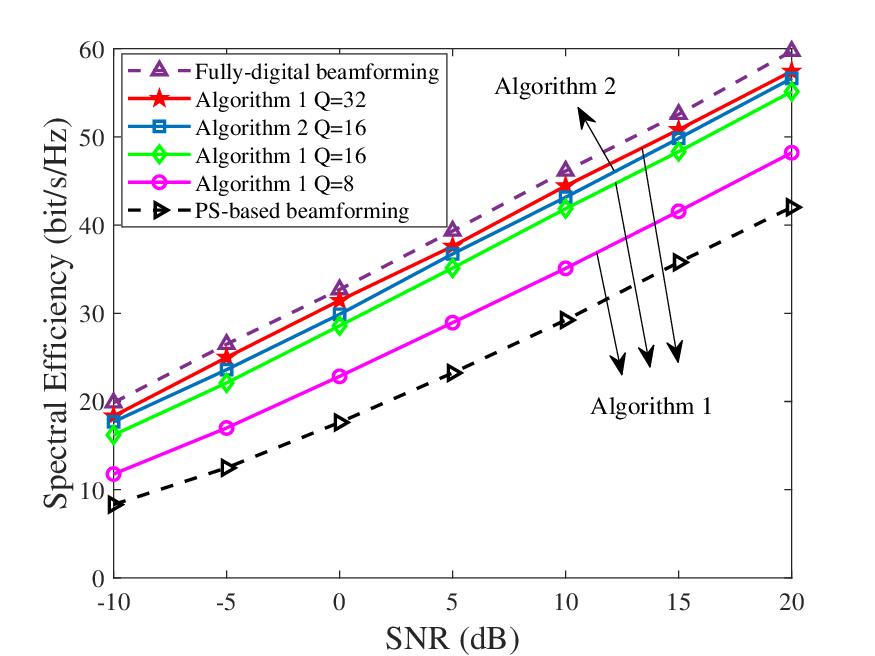} \vspace{-1.5mm}
\caption{The spectral efficiency varies with SNR for different beamforming schemes.}
\label{fig6}
\end{figure}

Fig. \ref{fig6} compares the spectral efficiency achieved by the TTD-based beamforming schemes proposed in \textbf{Algorithm \ref{algorithm1}} and \textbf{Algorithm \ref{algorithm2}} with that achieved by the fully-digital beamforming \cite{37} and the PS-based beamforming \cite{38}. As can be observed, the TTD-based beamforming schemes consistently outperform the PS-based beamforming scheme in terms of spectral efficiency. This is mainly because the TTD-based beamforming architecture takes advantage of TTD units to generate frequency-dependent phase shifts, thereby achieving more precise beamfocusing. This facilitates alleviating the beam squint effect and enhancing spectral efficiency performance. Additionally, as the number of $Q$ increases, \textbf{Algorithm \ref{algorithm1}} exhibits higher spectral efficiency and gradually approaches the optimal fully-digital beamforming performance. In particular, when SNR = 15 dB and $Q$ = 32, the spectral efficiency of \textbf{Algorithm \ref{algorithm1}} achieves 96.7\% of the performance of optimal fully-digital beamforming, whereas when $Q$ = 8, the spectral efficiency only reaches 80\%. Furthermore, it can also be observed that \textbf{Algorithm \ref{algorithm2}} demonstrates superior spectral efficiency performance compared to \textbf{Algorithm \ref{algorithm1}}. This is due to the dynamic adjustment of phase and delay parameters during the alternating optimization of PSs and TTDs, leading to a significant reduction in interference and an improvement in system performance.

\begin{figure}[t]
\centering
\vspace{-2mm}
\includegraphics[width=0.38\textwidth]{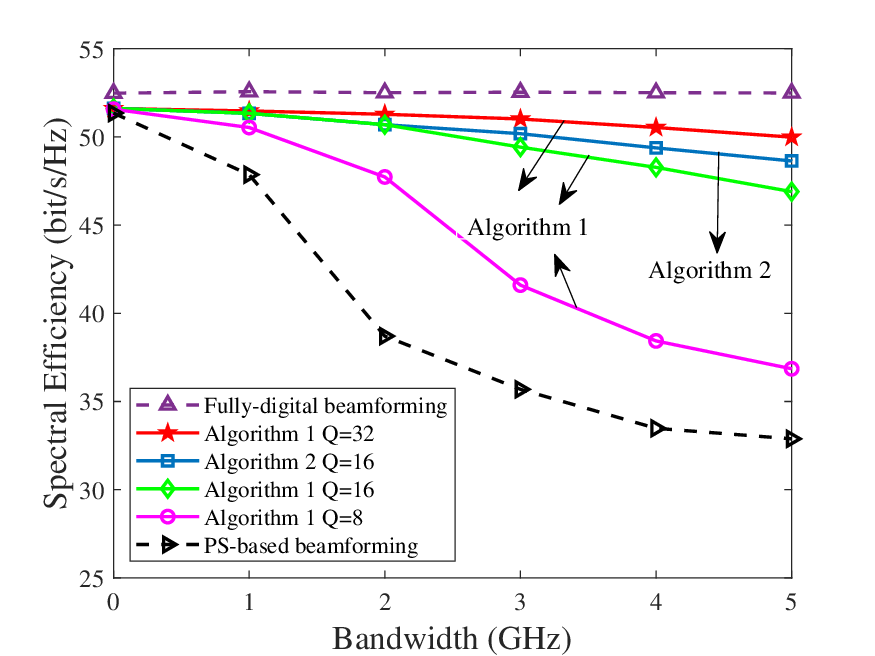}\vspace{-1mm}
\caption{The spectral efficiency varies with bandwidth for different beamforming schemes.}
\label{fig7}
\end{figure}

Fig. \ref{fig7} further investigates the relationship between the spectral efficiency and bandwidth at a fixed SNR of 15 dB. It is evident that as the bandwidth $B$ increases, the spectral efficiency of the PS-based beamforming scheme decreases significantly. This indicates that the spatial-frequency wideband effect has a negative impact on the system performance in conventional beamforming architectures. Compared to the PS-based beamforming scheme, the analytical beamforming scheme proposed in \textbf{Algorithm \ref{algorithm1}} and the joint-optimization beamforming scheme proposed in \textbf{Algorithm \ref{algorithm2}} demonstrate a slower decline in the spectral efficiency performance across the bandwidth range from $B$ = 0.1 GHz to $B$ = 5 GHz. To be specific, when the bandwidth is 3 GHz and $Q$ = 16, the PS-based beamforming scheme experiences a significant reduction of 31\% in spectral efficiency performance, whereas \textbf{Algorithm \ref{algorithm2}} and \textbf{Algorithm \ref{algorithm1}} only decrease by 4.5\% and 5.9\%, respectively. Moreover, Fig. \ref{fig7} also exhibits that the downtrend of spectral efficiency becomes slow as the number of $Q$ increases. These phenomena confirm the effectiveness of TTD-based beamforming schemes in mitigating spatial-frequency wideband effect. Consequently, we conclude that the TTD-based beamforming schemes are more applicable for larger bandwidth scenarios.

\begin{figure}[t]
\centering
\vspace{-1mm}
\includegraphics[width=0.355\textwidth]{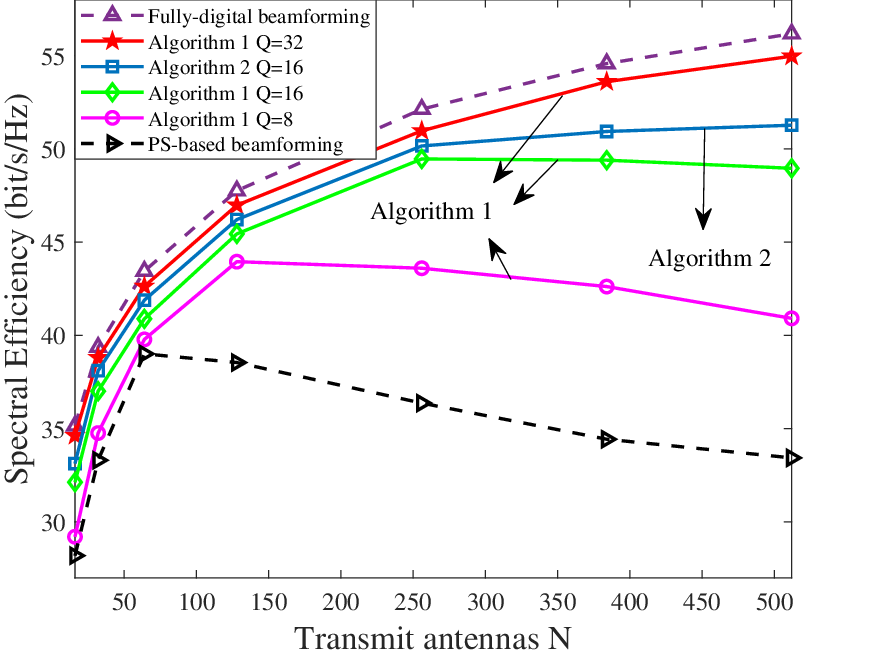}\vspace{-1mm}
\caption{The spectral efficiency varies with the number of transmit antenna $N$ for different beamforming schemes.}
\label{fig8}
\end{figure}

In Fig. \ref{fig8}, the impact of the number of transmit antennas $N$ on the spectral efficiency is investigated. Specifically, the number of antennas is gradually increased from 16 to 512. It can be seen that as the number of antenna elements increases, the PS-based beamforming scheme suffers from a severe spectral efficiency loss compared to the optimal fully-digital beamforming scheme. On the contrary, the proposed \textbf{Algorithm \ref{algorithm1}} with $Q$ = 32 exhibits only a slight performance loss. This indicates the effectiveness of TTD-based beamforming schemes in alleviating the near-field beam squint effect. Furthermore, for $Q$ = 16, both \textbf{Algorithm \ref{algorithm2}} and \textbf{Algorithm \ref{algorithm1}} demonstrate that the spectral efficiency performance shows an initial improvement followed by a subsequent decline as the number of transmit antennas increases. This is mainly because the relatively small number of TTD units may not be sufficient to achieve precise beamfocusing in XLAAs system, resulting in a decrease in spectral efficiency. Therefore, it is essential to determine the optimal number of TTD units based on our analysis in Section \ref{sectionIV} to maximize system performance while reducing hardware complexity.

Fig. \ref{fig9} depicts the effect of the maximum time delay $\tau_{\max }$ of TTD units on the spectral efficiency, with a fixed number of 32 TTD units. It is evident that the spectral efficiency performances of the TTD-based beamforming schemes are significantly affected by the value of $\tau_{\max}$. In particular, as the time delay value gradually increases, both \textbf{Algorithm \ref{algorithm1}} and \textbf{Algorithm \ref{algorithm2}} exhibit a significant improvement in spectral efficiency. For instance, when the time delay value varies from 0 ns to 20 ns, \textbf{Algorithm \ref{algorithm2}} and \textbf{Algorithm \ref{algorithm1}} demonstrate a 41\% and 37\% improvement in spectral efficiency, respectively. Remarkably, both \textbf{Algorithm \ref{algorithm1}} and \textbf{Algorithm \ref{algorithm2}} eventually reach a plateau with only moderate degradation relative to the optimal fully-digital solution. The spectral efficiency stabilizes after a certain point. This phenomenon highlights that the time delay values are adjusted to the optimal value to achieve the optimal beamfocusing in the desired user directions. Therefore, it is necessary to select an appropriate time delay value since excessive values do not provide any benefits and only contribute to unnecessary delays.
\begin{figure*}[b]
\small
\hrulefill 
\begin{align} \label{41}
\begin{array}{l}
{G_m}\left( {r,{\phi_1 },r,{\phi_2}} \right) = \frac{1}{N}\left| {{e^{jr({k_c} - {k_m})}}\sum\limits_{n = 0}^{N - 1} {{e^{j{k_m}\sqrt {{r^2} + {R^2} - 2rR\cos \left( {\phi_2   - {\psi _n}} \right)} }}} } \right.\left. { \times {e^{ - j{k_c}\sqrt {{r^2} + {R^2} - 2rR\cos \left( {{\phi_1} - {\psi _n}} \right)} }}} \right|\\
\hspace{22.2mm}\mathop  \approx \limits^{(a)} \frac{1}{N}\left| {{e^{jr({k_c} - {k_m})}}\sum\limits_{n = 0}^{N - 1} {{e^{j{k_m}\left[ {r - R\cos \left( {\phi_2 - {\psi _n}} \right)} \right]}}} } \right.\left. { \times {e^{ - j{k_c}\left[ {r - R\cos \left( {{\phi_1} - {\psi _n}} \right)} \right]}}} \right| = \frac{1}{N}\left| {\sum\limits_{n = 0}^{N - 1} {{e^{-jR\left[ {{k_m}\cos \left( {{\phi_2 } - {\psi _n}} \right) - {k_c}\cos \left( {\phi_1  - {\psi _n}} \right)} \right]}}} } \right|,
\end{array}
\tag{41} 
\end{align}
\begin{align} \label{50}
\begin{array}{l}
\hspace{-23mm}{G_m}\left( {{r_1},\phi ,{r_2},\phi } \right) \approx \frac{1}{N}\left| {\sum\limits_{n = 0}^{N - 1} {\left[ {\sum\limits_{{s_1} =  - \infty }^{ + \infty } {{j^{{s_1}}}{J_{{s_1}}}\left( {R\left( {{k_c} - {k_m}} \right)} \right){e^{j{s_1}\left( {\phi  - {\psi _n}} \right)}}} } \right] \times \left[ {\sum\limits_{{2s_2} =  - \infty }^{ + \infty } {{j^{{2s_2}}}{J_{{2s_2}}}\left( \varpi  \right){e^{2j{s_2}\left( {\phi  - {\psi _n}} \right)}}} } \right]} } \right|\\
\hspace{-0.5mm} \mathop  = \limits^{\left( d \right)} \frac{1}{N}\left| {\sum\limits_{{s_1} =  - \infty }^{ + \infty } {\sum\limits_{{2s_2} =  - \infty }^{ + \infty } {{j^{{s_1} + 2{s_2}}}{J_{{s_1}}}\left( {R\left( {{k_c} - {k_m}} \right)} \right){J_{{2s_2}}}\left( \varpi  \right) \times {e^{j \left( {{s_1} + 2{s_2}} \right) \phi }}\sum\limits_{n = 0}^{N - 1} {{e^{ - j\left( {{s_1} + 2{s_2}} \right){\psi _n}}}} } } } \right|,
\end{array}
\tag{50} 
\end{align}
\begin{equation}
\hspace{-65mm}{\sum\limits_{q = 0}^{Q-1} {{e^{ - j\left( {{k_c} - {k_m}} \right)R\cos \left( {{\phi _l} - {\theta _q}} \right)}}{e^{j\left( {{k_c} - {k_m}} \right)\frac{{{R^2}}}{{2{r_l}}}{{\sin }^2}\left( {{\phi _l} - {\theta _q}} \right)}}}  \times \sum\limits_{n = qP}^{\left( {q + 1} \right)P - 1} {{e^{ - j\left( {{s_1} + 2{s_2}} \right)\frac{{2\pi n}}{N}}}} }\nonumber\vspace{-6mm}
\end{equation}
\begin{align}
&\hspace{-1.5mm} \mathop  = \limits^{\left( g \right)} {e^{j\left( {{k_c} - {k_m}} \right)\frac{{{R^2}}}{{4{r_l}}}}}\sum\limits_{{s_3} =  - \infty }^{ + \infty } {\sum\limits_{2{s_4} =  - \infty }^{ + \infty }  {j^{{s_3} + 2{s_4}}}{J_{{s_3}}}\left( { - \left( {{k_c} - {k_m}} \right)R} \right){J_{2{s_4}}}\left( { - \left( {{k_c} - {k_m}} \right)\frac{{{R^2}}}{{4{r_l}}}} \right){e^{j\left( {{s_3} + 2{s_4}} \right){\phi _l}}}\frac{{1 - {e^{ - j\left( {{s_1} + 2{s_2}} \right)2\pi /Q}}}}{{1 - {e^{ - j\left( {{s_1} + 2{s_2}} \right)2\pi /N}}}}{e^{ - j\left( {{s_3} + 2{s_4}} \right)\frac{\vartheta }{Q}}}}\nonumber \\
&\hspace{+2mm} \times \sum\limits_{q = 0}^{Q-1} {e^{ - j\left( {{s_1} + 2{s_2} + {s_3} + 2{s_4}} \right)2\pi q/Q}}
\tag{54} 
\label{54}
\end{align}
\end{figure*}

\section{\normalsize{C}\footnotesize{ONCLUSION}}\label{sectionVII}\vspace{-0.5mm}
This paper investigated the beamforming optimization problem in near-field wideband UCA systems. Specifically, the beamfocusing property in both the distance and angular domain was characterized, revealing that the conventional beamforming architecture faced severe beamforming loss in near-field wideband UCA systems. To achieve an ideal beamforming gain, the analytical beamforming algorithm and the joint-optimization beamforming algorithm were proposed. Simulation results demonstrated the effectiveness of the proposed beamforming schemes in mitigating the beam squint effect and facilitating the beamfocusing in wideband UCA systems. For future works, we will further explore the channel estimation algorithms and low-complexity beam training methods for wideband UCA systems in near-field communications.
\begin{figure}[t]
\centering
\vspace{-1.5mm}
\includegraphics[width=0.383\textwidth]{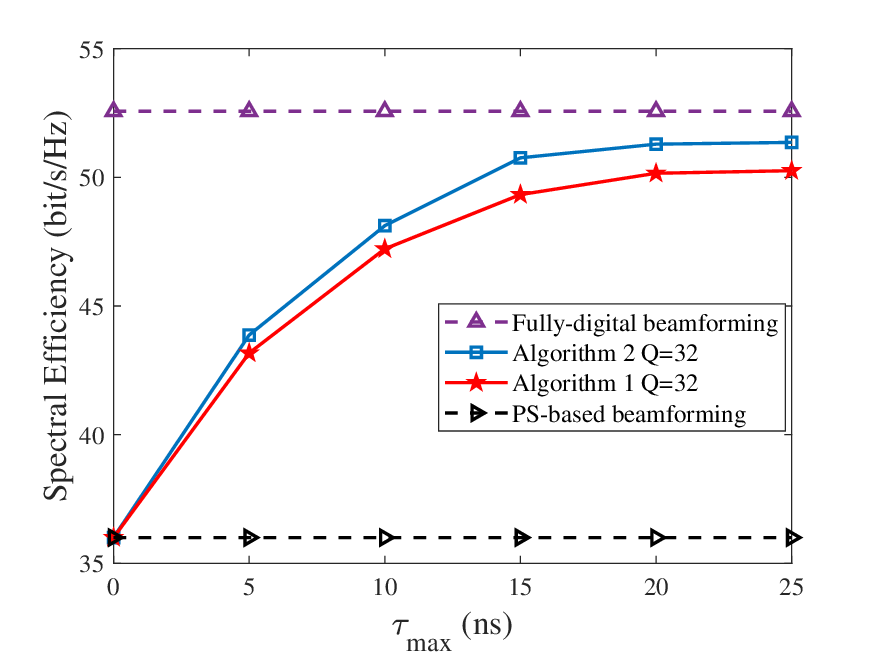}\vspace{-1.5mm}
\caption{The spectral efficiency varies with the maximum time delay $\tau_{\max }$ of TTD units.}
\label{fig9}
\end{figure}

\begin{appendices}
\section{Proof of \textbf{Lemma \ref{lemma_1}}} \label{proofA}\vspace{-0.5mm}
With the assumption ${r_1} = {r_2} = r$, the expression for beamforming gain \eqref{7} can be further represented as \eqref{41}, where ${\left( a \right)}$ stems from the first-order of Taylor series expansion. Then, we utilize the expansion of the Bessel function as \cite{32} \setcounter{equation}{41}\vspace{-1.5mm}
\begin{equation}
\label{42}
{e^{js\cos \phi }} = \sum\limits_{m =  - \infty }^{ + \infty } {{j^m}{J_m}\left( s \right){e^{jm\phi }}},\vspace{-1.5mm}
\end{equation}
where ${J_m}\left(  \cdot  \right)$ denotes the $m$-order Bessel function. Consequently, the beamforming gain can be rewritten as\vspace{-1mm}
\begin{equation}
\label{43}
\begin{array}{l}
\hspace{-2mm}{G_m}\!\left({r,\phi_1 ,r,{\phi_2}} \!\right) \!=\! \frac{1}{N}\!\!\left| {\sum\limits_{{s_1} =  - \infty }^{ + \infty } \!{\sum\limits_{{s_2} =  - \infty }^{ + \infty } {{j^{{s_1} + {s_2}}}{J_{{s_1}}}\!\!\left( { R{k_c}} \right)\!{J_{{s_2}}}\!\!\left( {- R{k_m}} \right)} } } \right.\\
\hspace{30mm}\left. { \times {e^{j\left( {{s_1}\phi_1  + {s_2}{\phi _2}} \right)}}\sum\limits_{n = 0}^{N - 1} {{e^{ - j\left( {{s_1} + {s_2}} \right){\psi _n}}}} } \right|.\vspace{-1mm}
\end{array}
\end{equation}

Recalling ${\psi _n} = \frac{{2\pi n}}{N}$, we can represent the last summation over $n$ as a piecewise function,
\begin{equation}
\label{44}
\sum_{n=0}^{N-1} e^{-j\left(s_1+s_2\right) \psi_n}=\left\{\begin{aligned}
N,\hspace{1mm} & s_1+s_2=N \cdot q, q \in \mathbb{Z} \\
0,\hspace{1mm} & s_1+s_2 \neq N \cdot q, q \in \mathbb{Z}.
\end{aligned}\right.
\end{equation}

Note that the piecewise function highlights the periodicity of the summation, indicating that the summation is zero unless ${{\rm{s}}_1} + {{\rm{s}}_2}$ is an integer multiple of the antenna number $N$. Next, we make the assumption that the number of transmit antenna elements $N$ is sufficiently large, which is a commonly adopted assumption in UCA analysis. Subsequently, we utilize the asymptotic property of the Bessel function, i.e.,\vspace{-1mm}
\begin{equation}
\label{45}
\left| {{J_{\left| {{s_1} + {s_2}} \right|}}\left( x \right)} \right| \le {\left( {\frac{{xe}}{{2\left| {{s_1} + {s_2}} \right|}}} \right)^{\left| {{s_1} + {s_2}} \right|}},
\end{equation}
where the value of $\left| {{J_{\left| {{s_1} + {s_2}} \right|}}\left( x \right)} \right|$ becomes negligible for large ${{s_1} + {s_2}}$. Consequently, we can conclude that $\left| {{J_{\left| {{s_1} + {s_2}} \right|}}\left( x \right)} \right| \approx 0$ could be assumed for ${s_1} + {s_2} = N \cdot q$ with $q \ne 0$ and large $N$. In other words, when assuming a large value for $N$, the conditions of ${s_1} + {s_2} \ne 0$ can be omitted in the summation. By substituting ${s_1} + {s_2} = 0$, the beamforming gain in \eqref{43} could be approximated as
\begin{equation}
\label{46}
\begin{array}{l}
\hspace{-2mm} {G_m}\! \left( {r, {\phi_1 }, r, {\phi_2}} \right) \mathop  = \limits^{(b)}\left| {\sum\limits_{s =  - \infty }^{ + \infty } {{J_s}\left( {R{k_c}} \right){J_s}\left( {R{k_m}} \right){e^{js\left( {{\phi_1 } - \phi_2 } \right)}}} } \right|,
\end{array}
\end{equation}
where ${\left( b \right)}$ is obtained by utilizing the property ${J_s}\left( { - x} \right) = {\left( { - 1} \right)^s}{J_s}\left( x \right)$ and ${J_{ - s}}\left( x \right) = {\left( { - 1} \right)^s}{J_s}\left( x \right)$. According to the addition theorems of Bessel functions \cite{32}
\begin{equation}
\label{47}
{J_0}\left( {{r_0}} \right) = \sum\limits_{s =  - \infty }^{ + \infty } {{J_s}\left( {{r_1}} \right){J_s}\left( {{r_2}} \right){e^{js\theta }}},
\end{equation}
where ${r_0} = \sqrt {r_1^2 + r_2^2 - 2{r_1}{r_2}\cos \theta } $, the beamforming gain could be further simplified as
\begin{equation}
\label{48}
{G_m}\left( {r,{\phi_1},r,{\phi_2}} \right) = \left| {{J_0}\left( \eta  \right)} \right|,
\end{equation}
where $\eta  = R\sqrt {k_c^2 + k_m^2 - 2{k_c}{k_m}\cos \left( {{\phi_1} - {\phi_2}} \right)} $. This completes the proof of \textbf{Lemma \ref{lemma_1}}.

\vspace{-4mm}
\section{Proof of \textbf{Lemma \ref{lemma_2}}} \label{proofB}
With the condition ${\phi _1} = {\phi _2} = \phi $, we obtain
\begin{equation}
\label{49}
\begin{array}{l}
{G_m}\left( {{r_1},\phi ,{r_2},\phi } \right) \mathop  = \limits^{(c)} \frac{1}{N}\left| {\sum\limits_{n = 0}^{N - 1} {{e^{jR\left( {{k_c} - {k_m}} \right)\cos \left( {\phi  - {\psi _n}} \right)}}} } \right.\\
\hspace{29mm}\left. { \times {e^{j{R^2}\left( {\frac{{{k_m}}}{{4{r_1}}} - \frac{{{k_c}}}{{4{r_2}}}} \right)}}{e^{j\varpi \cos \left( {2\phi  - 2{\psi _n}} \right)}}} \right|,
\end{array}
\end{equation}
where $\varpi  = {R^2}\left( { - \frac{{{k_m}}}{{4{r_1}}} + \frac{{{k_c}}}{{4{r_2}}}} \right)$ and equation ${\left( c \right)}$ is obtained by utilizing trigonometric functions $1 - {\cos ^2}\left( x \right) = \frac{{1 - \cos \left( {2x} \right)}}{2}$.

Based on the expansion of equation \eqref{42}, the beamforming gain can be derived as \eqref{50} at the bottom of this page, where ${\left( d \right)}$ is derived by exchanging the order of summation. Similar to the derivation process of Appendix \ref{proofA}, the conditions of $s_1 + 2s_2 \ne 0$  could be omitted in the summation when assuming a large transmit antenna elements $N$. By substituting $s_1 + 2s_2$ = 0, the beamforming gain is given by\vspace{-1mm}\setcounter{equation}{50}
\begin{equation}
\label{51}
\begin{array}{l}
\hspace{-2mm}{G_m}\left( {{r_1},\phi ,{r_2},\phi } \right) = {\left| {\sum\limits_{s =  - \infty }^{ + \infty }  {J_s}\left( {R\left( {{k_c} - {k_m}} \right)} \right){J_{ - s}}\left( \varpi  \right)} \right|}\vspace{+1mm}\\
\hspace{22.4mm} {\mathop  = \limits^{\left( e \right)} } \left| {{J_0}\left( {R\left( {{k_c} - {k_m}} \right) + \varpi } \right)} \right|,
\end{array}
\end{equation}
where ${\left( e \right)}$ utilizes the summation theorems of Bessel functions, i.e., \hspace{-2mm}$\sum\limits_{m =  - \infty }^\infty \hspace{-2mm} {{J_m}\left( z \right)\hspace{-0.7mm}{J_{n - m}}\left( t \right)}  \hspace{-0.7mm}=\hspace{-0.7mm}{J_n}\!\left({z + t} \right)$. \hspace{-0.3mm}This \hspace{-0.3mm}completes \hspace{-0.3mm}the \hspace{-0.3mm}proof.

\vspace{-6.5mm}
\section{Proof of \textbf{Lemma \ref{lemma_3}}} \label{proofC}\vspace{-1mm}
Combining \eqref{19} with \eqref{17}, the normalized array gain can be written as\vspace{+1mm}
\begin{equation}
\hspace{-53mm}{G_m}\left( {{\bf{W}}_{l,m}^{{\rm{TTD}}},{r_l},{\phi _l}} \right)\nonumber \vspace{-4mm}
\end{equation}
\begin{align}
\label{52}
=\!\! \frac{1}{N}\!\!\sum\limits_{q = 0}^{Q - 1} \!\!{\left[ {\sum\limits_{n = qP}^{\left( {q + 1} \right)P - 1} \!\!{{e^{j\frac{{2\pi }}{c}\left( {{f_c} - {f_m}} \right)\!\left[ {R\cos \left( {{\phi _l} - \frac{{2\pi n}}{N}} \right) - \frac{{{R^2}}}{{2{r_l}}}{{\sin }^2}\left( {{\phi _l} - \frac{{2\pi n}}{N}} \right)} \right]}}} } \right.}\nonumber \\
 &\hspace{-74mm} \left. { \times {e^{ - j\frac{{2\pi }}{c}\left( {{f_c} - {f_m}} \right)\left[ {R\cos \left( {{\phi _l} - {\theta _q}} \right) - \frac{{{R^2}}}{{2{r_l}}}{{\sin }^2}\left( {{\phi _l} - {\theta _q}} \right)} \right]}}} \right].
\end{align}

Then, by utilizing equation \eqref{42}, we can express the inner summation over $n$ as
\begin{equation}
\hspace{-11mm}{\sum\limits_{n = qP}^{\left( {q + 1} \right)P - 1} {{e^{j\left( {{k_c} - {k_m}} \right)\left[ {R\cos \left( {{\phi _l} - \frac{{2\pi n}}{N}} \right) - \frac{{{R^2}}}{{2{r_l}}}{{\sin }^2}\left( {{\phi _l} - \frac{{2\pi n}}{N}} \right)} \right]}}} }\nonumber\vspace{-6mm}
\end{equation}
\begin{align}
\label{53}
\hspace{-2.5mm}\mathop  = \limits^{\left( f \right)} \! {e^{ - j\left( {{k_c} - {k_m}} \right)\frac{{{R^2}}}{{4{r_l}}}}}\!\!\!\sum\limits_{n = qP}^{\left( {q + 1} \right)\!P - 1} \!\!\!{{e^{j\left( {{k_c} - {k_m}} \right)\! \left[\!{R\!\cos \left(\! {{\phi _l} - \frac{{2\pi n}}{N}} \!\right) + \frac{{{R^2}}}{{4{r_l}}}\!\cos \left(\! {2{\phi _l} \!-\! \frac{{4\pi n}}{N}} \right)} \!\right]}}}\nonumber \\
&\hspace{-91.4mm} = \!{{e^{ - j\left( {{k_c} - {k_m}} \right)\frac{{{R^2}}}{{4{r_l}}}}}\sum\limits_{{s_1} =  - \infty }^{ + \infty }  \sum\limits_{-2{s_2} =  - \infty }^{ + \infty }  {j^{{s_1} - 2{s_2}}}{J_{{s_1}}}\left( {\left( {{k_c} - {k_m}} \right)R} \right)}\nonumber\\
{ \times {J_{-2{s_2}}}\!\!\left( \!\!-{\left( {{k_c} \!-\! {k_m}} \right)\frac{{{R^2}}}{{4{r_l}}}} \right)\!{e^{j\left( {{s_1} - 2{s_2}} \right){\phi _l}}}\!\!\sum\limits_{n = qP}^{\left( {q + 1} \right)P - 1} \!\! {e^{- j\left( {{s_1} - 2{s_2}} \right)\frac{{2\pi n}}{N}}}},
\end{align}
where ${\left( f \right)}$ stems from the trigonometric functions, i.e., ${\sin ^2}\left( x \right) = \frac{{1 - \cos \left( {2x} \right)}}{2}$.

Furthermore, the outer summation over $q$ can be alternatively represented as equation \eqref{54} at the bottom of page 12. Equation ${\left( g \right)}$ is obtained by setting $\frac{\vartheta }{Q} = {\theta _q} - \frac{{2\pi q}}{Q}$. Assuming a relatively large $Q$, we can employ a similar approach to the proof of Appendix \ref{proofA} by setting ${s_1} - 2{s_2} =0$  and ${s_3} - 2{s_4} = 0$. Consequently, the beamforming gain can be rewritten as \setcounter{equation}{54}
\begin{equation}
\hspace{-53mm} {G_m}\left({{\bf{W}}_{l,m}^{{\rm{TTD}}},{r_l},{\phi _l}} \right)\nonumber\vspace{-4mm}
\end{equation}
\begin{align}
\label{55}
\mathop  =  \frac{1}{P}\sum\limits_{{s_1} =  - \infty }^{ + \infty } {\sum\limits_{2{s_2} =  - \infty }^{ + \infty } {J_{{s_1} - 2{s_2}}^2\left( {\left( {{k_c} - {k_m}} \right)R - \left( {{k_c} - {k_m}} \right)\frac{{{R^2}}}{{4{r_l}}}} \right)} } \nonumber\\
&\hspace{-87mm}\times \sum\limits_{j = 0}^{P - 1} {{e^{ - j\left( {{s_1} -2{s_2}} \right)\frac{{2\pi j - P\vartheta }}{N}}}}\nonumber\\
&\hspace{-90.8mm} \mathop  = \limits^{\left( h \right)} \frac{1}{P}\sum\limits_{j = 0}^{P - 1} {{J_0}\left( {{R_j}} \right)},
\end{align}
where ${{R_j} = \sqrt 2 \left( {{k_c} - {k_m}} \right)\!R\!\left( {1 \!\!-\!\! \frac{R}{{4r_l}}} \right)\!\!\sqrt {1 \!-\! \cos \!\left( {\frac{{2\pi j }}{N} \!-\! \frac{\vartheta  }{Q}} \right)} }$ and ${\left( h \right)}$ is derived from the property of the addition theorems of Bessel function. It can be observed from equation \eqref{55} that the Bessel function reaches its maximum value when $\vartheta  = \pi  - \frac{\pi }{P}$. Consequently, the optimal beamforming gain design of frequency-dependent ${\bf{W}}_{l,m}^{{\rm{TTD}}}$ can be formulated as
\begin{equation}
\label{56}
{\left( {{\bf{W}}_{l,m}^{{\rm{TTD}}}} \right)_q}= {e^{j\left( {{k_c} - {k_m}} \right)\left[ {{r_l} - R\cos \left( {{\phi _l} - {\theta _q}} \right) + \frac{{{R^2}}}{{2{r_l}}}{{\sin }^2}\left( {{\phi _l} - {\theta _q}} \right)} \right]}},
\end{equation}
where ${\theta _q} = \frac{{\left( {P - 1} \right)\pi }}{N} + \frac{{2\pi q}}{Q}$. This completes the proof.

\end{appendices}


\begin{thebibliography}{}

\end{thebibliography}


\vspace{-1mm}
\begin{thebibliography}{99}\vspace{-1mm}
\bibitem{1} E. C. Strinati, S. Barbarossa, J. L. Gonzalez-Jimenez, D. Ktenas, N. Cassiau, and L. Maret, ``6G: The next frontier: From holographic messaging to artificial intelligence using subterahertz and visible light communication,'' \emph{IEEE Veh. Technol. Mag.}, vol. 14, no. 3, pp. 42-50, Sept. 2019.

\bibitem{2} B. Ji, Y. Han, S. Liu, F. Tao, G. Zhang, Z. Fu, and C. Li, ``Several key technologies for 6G: Challenges and opportunities,'' \emph{IEEE Commun. Mag.}, vol. 5, no. 2, pp. 44-51, Jun. 2021.

\bibitem{3} I. F. Akyildiz, C. Han, Z. Hu, S. Nie, and J. M. Jornet, ``Terahertz band communication: An old problem revisited and research directions for the next decade,'' \emph{IEEE Trans. Commun.}, vol. 70, no. 6, pp. 4250-4285, Jun. 2022.

\bibitem{4} H.-J. Song and N. Lee, ``Terahertz communications: Challenges in the next decade,'' \emph{IEEE Trans. THz Sci. Technol.}, vol. 12, no. 2, pp. 105-117, Mar. 2022.

\bibitem{5} H. Lu and Y. Zeng, ``How does performance scale with antenna number for extremely large-scale MIMO?'' in \emph{Proc. IEEE Int. Conf. Commun.}, Montreal, QC, Canada, Aug. 2021, pp. 1-6.

\bibitem{6} H. Lu and Y. Zeng, ``Near-field modeling and performance analysis for multi-user extremely large-scale MIMO communication,'' \emph{IEEE Commun. Lett.}, vol. 26, no. 2, pp. 277-281, Feb. 2022

\bibitem{7} M. Cui, Z. Wu, Y. Lu, X. Wei, and L. Dai, ``Near-field MIMO communications for 6G: Fundamentals, challenges, potentials, and future directions,'' \emph{IEEE Commun. Mag.}, vol. 61, no. 1, pp. 40-46, Jan. 2023.

\bibitem{8} K. T. Selvan and R. Janaswamy, ``Fraunhofer and Fresnel distances: Unified derivation for aperture antennas,'' \emph{IEEE Antennas Propag. Mag.}, vol. 59, no. 4, pp. 12-15, Aug. 2017.

\bibitem{9} E. Bjornson, O. T. Demir, and L. Sanguinetti, ``A primer on near-field beamforming for arrays and reconfigurable intelligent surfaces,'' in \emph{Proc. 55th Asilomar Conf. Signals, Syst., Comput.}, Pacific Grove, CA, USA, Oct. 2021, pp. 105-112.

\bibitem{10} Y. Liu, Z. Wang, J. Xu, C. Ouyang, X. Mu, and R. Schober, ``Near-field communications: A tutorial review,'' \emph{IEEE Open J. Commun. Soc.}, vol. 4, pp. 1999-2049, Aug. 2023.

\bibitem{11} H. Zhang, N. Shlezinger, F. Guidi, D. Dardari, M. F. Imani, and Y. C. Eldar, ``Beam focusing for near-field multiuser MIMO communications,'' \emph{IEEE Trans. Wireless Commun.}, vol. 21, no. 9, pp. 7476-7490, Mar. 2022.

\bibitem{12} R. Ji, S. Chen, C. Huang, J. Yang, W. E. I. Sha, Z. Zhang, C. Yu, and M. Debbah, ``Extra DoF of near-field holographic MIMO communications leveraging evanescent waves,'' \emph{IEEE Wireless Commun. Lett.}, vol. 12, no. 4, pp. 1-5, Apr. 2023.

\bibitem{13} Z. Zhang and L. Dai, ``Pattern-division multiplexing for continuous-aperture MIMO,'' in \emph{Proc. IEEE Int. Conf. Commun.}, Seoul, Republic of Korea, May. 2022, pp. 3287-3292.

\bibitem{14} H. Lu and Y. Zeng, ``Communicating with extremely large-scale array/surface: Unified modeling and performance analysis,'' \emph{IEEE Trans. Wireless Commun.}, vol. 21, no. 6, pp. 4039-4053, Jun. 2022.

\bibitem{15} C. Han, L. Yan, and J. Yuan, ``Hybrid beamforming for terahertz wireless communications: Challenges, architectures, and open problems,'' \emph{IEEE Wireless Commun.}, vol. 28, no. 4, pp. 198-204, Aug. 2021.

\bibitem{16} Y. Gao, M. Khaliel, F. Zheng, and T. Kaiser, ``Rotman lens based hybrid analog-digital beamforming in massive MIMO systems: Array architectures, beam selection algorithms and experiments,'' \emph{IEEE Trans. Veh. Technol.}, vol. 66, no. 10, pp. 9134-9148, Oct. 2017.

\bibitem{17} H. Zhang, H. Zhang, W. Liu, K. Long, J. Dong, and V. C. M. Leung, ``Energy efficient user clustering, hybrid precoding and power optimization in terahertz MIMO-NOMA systems,'' \emph{IEEE J. Sel. Areas Commun.}, vol. 38, no. 9, pp. 2074-2085, Sep. 2020.

\bibitem{18} F. Gao, L. Xu, and S. Ma, ``Integrated sensing and communications with joint beam-squint and beam-split for mmWave/THz massive MIMO,'' \emph{IEEE Trans. Commun.}, vol. 71, no. 5, pp. 2963-2976, May. 2023.

\bibitem{19} R. Zhang, W. Hao, G. Sun, and S. Yang, ``Hybrid precoding design for wideband THz massive MIMO-OFDM systems with beam squint,'' \emph{IEEE Syst. J.}, vol. 15, no. 3, pp. 3925-3928, Sept. 2021.

\bibitem{20} M. Cui and L. Dai, ``Near-field wideband beamforming for extremely large antenna arrays,'' \emph{arXiv preprint arXiv:2109.10054}, Dec. 2023.

\bibitem{21} Z. Wang, X. Mu, and Y. Liu, ``Beamfocusing optimization for near-field wideband multi-user communications,'' \emph{arXiv preprint arXiv: 2306.16861}, Nov. 2023.

\bibitem{22} Y. Zhang and A. Alkhateeb, ``Deep learning of near-field beam focusing in terahertz wideband massive MIMO systems,'' \emph{IEEE Wireless Commun. Lett.}, vol. 12, no. 3, pp. 535-539, Mar. 2023.

\bibitem{23} Y. Ji, W. Fan, and G. F. Pedersen, ``Near-field signal model for large-scale uniform circular array and its experimental validation,'' \emph{IEEE Antennas Wireless Propag. Lett.}, vol. 16, pp. 1237-1240, Nov. 2017.

\bibitem{24} M. Cui, L. Dai, Z. Wang, S. Zhou, and N. Ge, ``Near-field rainbow: Wideband beam training for XL-MIMO,'' \emph{IEEE Trans. Wireless Commun.}, vol. 22, no. 6, pp. 3899-3912, Jun. 2023.

\bibitem{25} Y. Xie, B. Ning, L. Li, and Z. Chen, ``Near-field beam training in THz communications: the merits of uniform circular array,'' \emph{IEEE Wireless Commun. Lett.}, vol. 12, no. 4, pp. 575-579, Jan. 2023.

\bibitem{26} X. Cheng, Y. He, and J. Qiao, ``Channel modeling for UCA and URA massive MIMO systems,'' in \emph{Proc. Int. Conf. Comput., Netw. Commun.}, Big Island, HI, USA, Mar. 2020, pp. 963-968.

\bibitem{27} Z. Wu, M. Cui, and L. Dai, ``Enabling more users to benefit from near-field communications: From linear to circular array,'' \emph{IEEE Wireless Commun. Lett.}, Early Access DOI: 10.1109/TWC.2023.3310912.

\bibitem{28} Z. Zhou, X. Gao, J. Fang, and Z. Chen, ``Spherical wave channel and analysis for large linear array in LoS conditions,'' in \emph{Proc. IEEE Globecom Workshops}, San Diego, CA, USA, Dec. 2015, pp. 1-6.

\bibitem{29} J. Sherman, ``Properties of focused apertures in the Fresnel region,'' \emph{IRE Trans. Antennas Propag.}, vol. 10, no. 4, pp. 399-408, Jul. 1962.

\bibitem{30} B. Ning, Z. Tian, W. Mei, Z. Chen, C. Han, S. Li, J. Yuan, and R. Zhang, ``Beamforming technologies for ultra-massive MIMO in terahertz communications,'' \emph{IEEE Open J. Commun. Soc.}, vol. 4, pp. 614-658, Feb. 2023.

\bibitem{31} J. -J. Fuchs, ``On the application of the global matched filter to DOA estimation with uniform circular arrays,'' \emph{IEEE Trans. Signal Process.}, vol. 49, no. 4, pp. 702-709, Apr. 2001.

\bibitem{32} I.S. Gradshteyn and I.M. Ryzhik Notation, \emph{Table of Integrals, Series, and Products.} San Diego, California, USA: Elsevier Academic Press, 1985.

\bibitem{33} X. Zhang, Z. Wang, H. Zhang, and L. Yang, ``Near-field channel estimation for extremely large-scale array communications: A model-based deep learning approach,'' \emph{IEEE Commun. Lett.}, vol. 27, no. 4, pp. 1155-1159, Apr. 2023.

\bibitem{34} S. Tarboush, A. Ali, and T.Y. Al-Naffouri, ``Cross-Field channel estimation for ultra Massive-MIMO THz systems,'' \emph{IEEE Trans. Wireless Commun.}, Early Access DOI: 10.1109/TWC.2024.3352894.

\bibitem{35} E. Bjornson, M. Bengtsson, and B. Ottersten, ``Optimal multiuser transmit beamforming: A difficult problem with a simple solution structure,'' \emph{IEEE Signal Process. Mag.}, vol. 31, no. 4, pp. 142-148, Jul. 2014.

\bibitem{36} M. H. Ghazizadeh and A. Medi, ``Novel trombone topology for wideband true-time-delay implementation,'' \emph{IEEE Trans. Micro. Theory Tech.}, vol. 68, no. 4, pp. 1542-1552, Apr. 2020.

\bibitem{37} K. Shen and W. Yu, ``Fractional programming for communication systems-Part II: Uplink scheduling via matching,'' \emph{IEEE Trans. Signal Process.}, vol. 66, no. 10, pp. 2631-2644, May 2018.

\bibitem{38} A. Raafat, M. Sefun, A. Agustin, J. Vidal, E. A. Jorswieck, and Y. Corr, ``Energy efficient transmit-receive hybrid spatial modulation for large-scale MIMO systems,'' \emph{IEEE Trans. Commun.}, vol. 68, no. 3, pp. 1448-1463, Mar. 2020.

\end{thebibliography}
\end{document}